\documentclass[traditabstract]{aa} 
 
\usepackage{graphicx, times, txfonts}

\newcommand{\less}{\raisebox{-1.1mm}{$\stackrel{<}{\sim}$}} 
 
\newcommand{\msol}{\mbox{M$_{\odot}$}} 
 
\newcommand{\msolyr}{{M$_{\odot}$}\,yr$^{-1}$} 
\newcommand{\mdot}{$\dot{M}$}
\newcommand{\lsol}{\mbox{L$_{\odot}$}}

\newcommand{\ks}{km s$^{-1}$} 
 
\newcommand{\mum}{$\mu$m}

\begin{document}

\title{
Luminosities and mass-loss rates of SMC and LMC AGB stars and Red Supergiants
\thanks{
Tables~1-4 are available in electronic form at the CDS via 
anonymous ftp to cdsarc.u-strasbg.fr (130.79.128.5) or via 
http://cdsweb.u-strasbg.fr/cgi-bin/qcat?J/A+A/. 
Figures~1-3 and 5-7 are available in the on-line edition of A\&A. 
} 
}  
 
\author{ 
M.~A.~T.~Groenewegen 
\inst{1}  
\and 
G.~C.~Sloan
\inst{2} 
\and
I.~Soszy\'nski 
\inst{3}
\and
E.~A.~Petersen
\inst{4,5}
}

\institute{ 
Koninklijke Sterrenwacht van Belgi\"e, Ringlaan 3, B--1180 Brussels, Belgium \\ \email{marting@oma.be}
\and
Cornell University, Astronomy Department, Ithaca, NY 14853-6801, USA
\and
Warsaw University Observatory, Al. Ujazdowskie 4, 00-478 Warszawa, Poland
\and
University of Nebraska, Department of Physics and Astronomy, Lincoln, NE 68588,
USA
\and
NSF REU Research Assistant, Cornell University, Astronomy Department, Ithaca, NY 14853-6801, USA
} 
 
\date{received: 2009,  accepted: 20-08-2009} 
 
\offprints{Martin Groenewegen} 
 
 
\abstract {
{\it Context} Mass loss is one of the fundamental properties of Asymptotic Giant Branch (AGB) stars, and
through the enrichment of the interstellar medium, AGB stars are 
key players in the life cycle of dust and gas in the universe.
However, a quantitative understanding of the mass-loss process is 
still largely lacking, particularly its dependence on metallicity.

{\it Aims} To investigate the relation between mass loss, luminosity and
  pulsation period for a large sample of evolved stars in the Small
  and Large Magellanic Cloud.

{\it Results} Dust radiative transfer models are presented for 101 carbon stars and
  86 oxygen-rich evolved stars in the Magellanic Clouds for which 
  5--35~\mum\ {\it Spitzer} IRS spectra are available. The spectra are
  complemented with available optical and infrared photometry to
  construct the spectral energy distribution.  A minimisation procedure
  is used to fit luminosity, mass-loss rate and dust temperature at
  the inner radius.  Different effective temperatures and dust content
  are also considered.  Periods from the literature and from new
  OGLE-III data are compiled and derived.
We derive (dust) mass-loss rates and luminosities for the entire sample. 
Based on luminosities, periods and amplitudes and colours, the O-rich
stars are classified in foreground objects, AGB stars and Red Super Giants.
For the O-rich stars silicates based on laboratory optical constants
are compared to ``astronomical silicates''.  Overall, the grain type
by Volk \& Kwok (1988) fit the data best.  However, the fit based
on laboratory optical constants for the grains can be improved by
abandoning the small-particle limit.  The influence of grain size,
core-mantle grains and porosity are explored.
Relations between mass-loss rates and luminosity and pulsation period
are presented and compared to the predictions of evolutionary models,
those by Vassiliadis \& Wood (1993) and their adopted mass-loss recipe, 
and those based on a Reimers mass-loss law with a scaling of a factor 
of five.  The Vassiliadis \& Wood models describe the data better,
although there are also some deficiencies, in particular to the 
maximum adopted mass-loss rate.
The OGLE-III data reveal an O-rich star in the SMC with a period 
of 1749 days.  Its absolute magnitude of $M_{\rm bol}= -8.0$ makes it 
a good candidate for a super-AGB star.
}

\keywords{circumstellar matter -- infrared: stars -- 
stars: AGB and post-AGB -- stars: mass loss -- Magellanic Clouds} 

\maketitle

\section{Introduction} 

Almost all stars with initial masses in the range $\sim$ 0.9--8~\msol\ 
will pass through the asymptotic giant branch (AGB) phase, which is
the last stage of active nuclear burning before they become post-AGB
stars, planetary nebulae and finally white dwarfs.  Slightly more
massive stars will pass through the red supergiant (RSG) phase before
they may end as supernovae.
In both cases, mass-loss dominates the final evolutionary stages of the
star.

Although this is well-known and studied in detail in galactic sources
with the advent of the {\it Infrared Astronomical Satellite (IRAS)} and
the {\it Infrared Space Observatory (ISO)}, uncertainties in distances
lead to uncertainties in luminosities and mass-loss rates.  Sources at
known distances, as in Large and Small Magellanic Clouds (LMC and SMC),
reduce this problem, and also allow one to study the effect of 
metallicity on the mass-loss rate.

In a previous paper, Groenewegen et al. (2007) modelled the spectral
energy distribution (SED) and spectra taken with the Infrared
Spectrograph (IRS; Houck et al.\ 2004) onboard the {\it Spitzer Space 
Telescope} (Werner et al.\ 2004) for a sample of 60 carbon (C) stars.
They concluded that, assuming similar expansion velocities and
dust-to-gas ratios as in Galactic stars, mass-loss rates versus
luminosity or pulsation period scatter around the galactic relation
for sources in both the LMC and SMC sources.  In other words, there is
no evidence that the mass-loss rate of C stars depends on 
metallicity.
Recent theoretical work also suggests that lower metallicity does not
necessarily imply smaller mass-loss rates for carbon stars (Mattsson et
al.\ 2008; Wachter et al.\ 2008).  The detection of dust forming
around a C star in the Sculptor Dwarf Spheroidal galaxy (Sloan et
al. 2009), with a metallicity of [Z/H] = $-1.3$, supports the theory
observationally.

Sloan et al. (2008) compared {\it Spitzer} spectroscopy of C stars
and oxygen-rich AGB stars and RSGs (hereafter referred to as M stars).
They found that while the carbon stars showed little dependence of
mass loss on metallicity, the amount of dust produced by M stars
declined in more metal-poor environments.  The aim of the present
paper is to extend the analysis by considering nearly 90 M stars
and enlarging the sample of C stars to over 100.

Section~2 describes the sample of AGB stars and RSG with IRS spectra,
the photometry to be fitted, and the derivation of pulsation periods.
Section~3 describes the radiative transfer model and the properties
of the dust species considered.
Section~4 presents the results in the form of tables with the fitted
parameters and figures comparing the models to the SEDs and IRS spectra.
Section~5 discusses the results.  In particular, we attempt to
separate the O-rich stars into foreground, AGB stars and RSG. 
We discuss the influence of different assumptions on the shape and
size of the grain on the fit to the IRS spectra.  We also examine
how the mass-loss rate depends on luminosity and period and compare
our results to evolutionary models.
Section~6 summarises the findings.

\section{The sample} 

Several groups have obtained {\it Spitzer} IRS data of evolved stars 
in the LMC and SMC.
In this paper we consider the currently publically available data from 
the following programmes:
 200 (P.I.\ J.~Houck), 
3277 (P.I.\ M.~Egan), 
3426 (P.I.\ J.~Kastner), 
3505 (P.I.\ P.~Wood), and 
3591 (P.I.\ F.~Kemper).
The data in these programs are described by Sloan et al.\ (2008, program 
200), Sloan et al.\ (2006, program 3277), Buchanan et al.\ (2006, program 
3426), Zijlstra et al.\ (2006) and Lagadec et al.\ (2007) for program 
3505, and Leisenring et al.\ (2008, program 3591).
We have retrieved the spectra from these programs from the public
archive and reduced them in a uniform manner, as described by Sloan
et al.\ (2006, 2008).  The reader should refer to these works for
more details.  Here, we outline how spectra are produced from the
IRS data.  All of the spectra were taken
using the low-resolution modules on the IRS, Short-Low (SL) covering
5.2--14.3~\mum, and Long-Low (LL) covering 14.2--37.0~\mum.  The
standard low-resolution observation placed the source in two nod
positions in each spectral aperture.  The spectral images were 
differenced to remove background emission and cleaned to correct 
bad pixels.  Spectra were extracted from the images using the tools 
available in SPICE, which is distributed by the {\it Spitzer} 
Science Center.  Spectra from the separate nods were combined
using a spike-rejection algorithm to remove features in one nod
spectrum but not the other.  Spectra from the separate apertures
and modules were then combined, using multiplicative shifts to
remove discontinuities between them and finally removing extraneous
data from the ends of each spectral segment.

The five programs considered here did not exclusively observe AGB
stars and RSGs.
Targets were selected from these programs by examining the IRS spectra,
collecting additional photometry (see below), consulting SIMBAD and the
papers describing these programs, and considering our radiative transfer 
models (see Sect.~\ref{themodel}).  Excluded sources include those with 
very poor S/N IRS data,
sources where the SED and spectrum did not match at all (indicating that
the IRS peak-up was on a source other than the intended target),
sources with a likely disk geometry (invalidating the spherically 
symmetric radiative transfer model used here),
a post-AGB star (MSX SMC 029; Kraemer et al.\ 2006), 
two RCrB stars (MSX SMC 014, MSX SMC 155; Kraemer et al.\ 2005),
O/Be-stars, and
objects showing PAH emission\footnote{
In detail, the following sources from these programs are not considered.
From program 200: WBP 219, WBP 29, WBP51, NGC 371 LE8, NGC 381 LE31, NGC 371 LE28, WBP 104, WBP 116, HV 5810, HV 5680, IRAS 04530-6916; \\
from program 3277: MSX SMC 014, 029, 079, 125, 155, 180;  \\
from program 3426: MSX LMC 559, 1306, 1794, 217, 222, 22, 764, 836, 889, 894, 934, 773, 769, 890;  \\
from program 3505: NGC 419 LE 35, GM 106, RAW 1559, ISO-MCMS J005149.4-731315 (iso00518), IRAS 05328-6827; \\
from program 3591: MSX LMC 616, LHA 120-N 89, LHA 120-N 77d, BSDL 2894, BSDL 126, [O96] D010b-262, HD 38489, 
NGC 1978 WBT2665, NGC 1948 WBT2215, HD 269924, MSX LMC 795, MSX LMC 1786, [L63] 31, MSX LMC 906, 
[SL63] 482, IRAS 04514-6913, MSX LMC 610.
}. 
The remaining sample includes the very interesting object WOH G 64,
even though Ohnaka et al.\ (2008) have recently demonstrated with
interferometric observations that the mid-IR visibility curves and the 
SED can be better modelled with a torus.

The sample under consideration consists of 101 C-stars and 86 M-stars.
Tables~\ref{Tab-Csample} and \ref{Tab-Msample} 
lists basic information:  some common names (as listed by SIMBAD),
an OGLE-{\sc iii} identifier when this lightcurve is analysed and
shown in Figure~\ref{Fig-LC-O3}, R.A. and declination in decimal
degrees, the identifier used in figures and tables below, the adopted
pulsation period, the (semi-)amplitude of the adopted pulsation period
in the filter where the lightcurve was obtained, i.e.\ OGLE $I$, MACHO
$B,R$, ASAS $V,I$, or in the $K$-band 
(Only the first entries are shown for guidance; both tables are 
available in their complete form at the CDS).
The stars are listed in order of their luminosity, from brightest to
faintest as determined below (and assuming these sources are in the 
LMC and SMC).

\begin{figure*}[!ht]
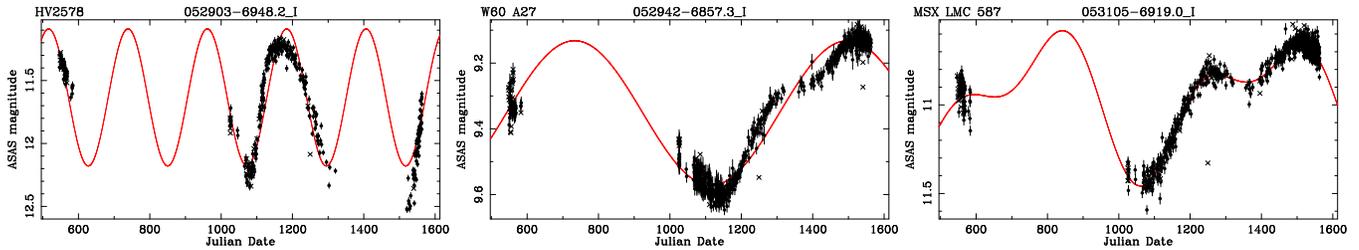


\begin{minipage}{0.32\textwidth}
\resizebox{\hsize}{!}{\includegraphics{052903-6948.2_I.ps}}
\end{minipage}
\begin{minipage}{0.32\textwidth}
\resizebox{\hsize}{!}{\includegraphics{052942-6857.3_I.ps}}
\end{minipage}
\begin{minipage}{0.32\textwidth}
\resizebox{\hsize}{!}{\includegraphics{053105-6919.0_I.ps}}
\end{minipage}

\caption[]{ 
Sample lightcurves and fits to ASAS data. 
The identifier used in the present paper and the ASAS identifier are 
listed on top of the plot.
The complete figure is available in the electronic edition. 
Julian Date plotted is JD-2450000. 
} 
\label{Fig-LC-ASAS} 
\end{figure*}

\begin{figure*}[!ht]
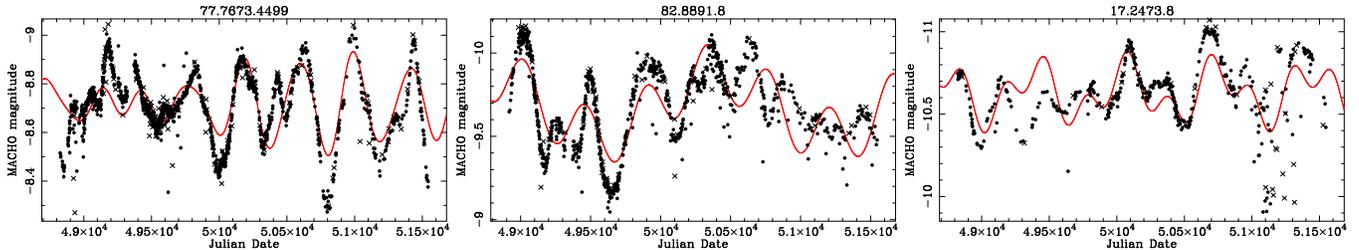


\begin{minipage}{0.32\textwidth}
\resizebox{\hsize}{!}{\includegraphics{77.7673.4499+++.ps}}
\end{minipage}
\begin{minipage}{0.32\textwidth}
\resizebox{\hsize}{!}{\includegraphics{82.8891.8++++++.ps}}
\end{minipage}
\begin{minipage}{0.32\textwidth}
\resizebox{\hsize}{!}{\includegraphics{17.2473.8++++++.ps}}
\end{minipage}

\caption[]{ 
Sample lightcurves and fits to MACHO data. 
MACHO identifiers are listed on top of the plot, and are cross-referenced in 
Tables~\ref{Tab-Csample} and \ref{Tab-Msample}. 
The complete figure is available in the electronic edition. 
Julian Date plotted is JD-2400000. 
} 
\label{Fig-LC-MACHO} 
\end{figure*}

\begin{figure*}[!ht]
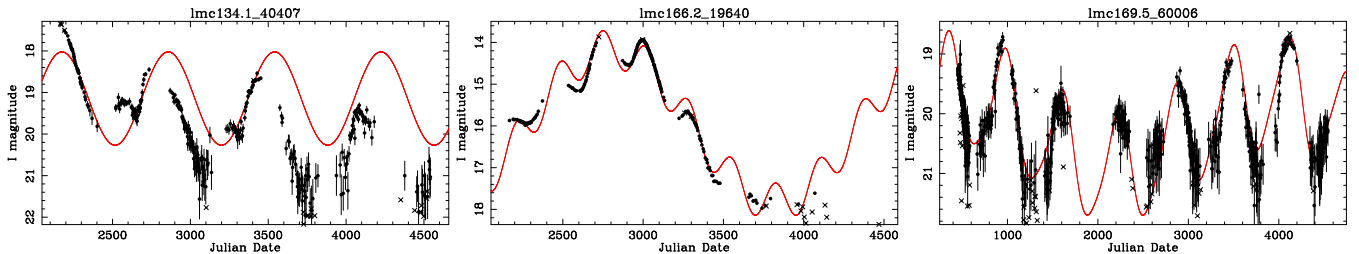


\begin{minipage}{0.32\textwidth}
\resizebox{\hsize}{!}{\includegraphics{lmc134.1_40407.ps}}
\end{minipage}
\begin{minipage}{0.32\textwidth}
\resizebox{\hsize}{!}{\includegraphics{lmc166.2_19640.ps}}
\end{minipage}
\begin{minipage}{0.32\textwidth}
\resizebox{\hsize}{!}{\includegraphics{lmc169.5_60006.ps}}
\end{minipage}

\caption[]{ 
Sample lightcurves and fits to OGLE-{\sc iii} data (and OGLE-{\sc ii} data 
when available). 
OGLE-{\sc iii} identifiers are listed on top of the plot, and are 
cross-referenced in Tables~\ref{Tab-Csample} and \ref{Tab-Msample}. 
The complete figure is available in the electronic edition. 
Julian Date plotted is JD-2450000. 
} 
\label{Fig-LC-O3} 
\end{figure*}

Many of the periods quoted in Tables~\ref{Tab-Csample} and \ref{Tab-Msample} 
come from the published literature, but in some cases publically
available data were re-analysed if the quoted period did not seem to
match the lightcurve or if published periods did not agree with each other.
Figures~\ref{Fig-LC-ASAS} and ~\ref{Fig-LC-MACHO} show 
our fits to publically availably
ASAS\footnote{http://www.astrouw.edu.pl/asas/} (Pojmanski 2002) and
MACHO data\footnote{see http://wwwmacho.anu.edu.au/}.
In addition, we used unpublished data from the OGLE-{\sc iii} survey
(Udalski et al.\ 2008), and when available, combined this with 
OGLE-{\sc ii} data.  In these cases, Figure~\ref{Fig-LC-O3}
shows the observed data and the fitted lightcurve.
Tables~\ref{Tab-Csample} and \ref{Tab-Msample} list the adopted
pulsation period, but many of the stars for which we (re-)analysed the
light curve show additional periods. These are listed in Appendix~A.


\begin{table*}

  \caption{The C-star sample: identifiers and pulsation periods for 
  the first few entries.}
  \begin{tabular}{lrrlrrr}
  \hline
Names &  R.A.      & Declination & Identifier & Period & Ref. & Ampl. (Filter) \\
\hline

IRAS 04496-6958, MSX LMC 1130          &  72.327000  & -69.887361  &   iras04496  & 723, 741 & 1, 18 & 0.44 ($K$) \\ 
IRAS 05278-6942, MSX LMC  635          &  81.850458  & -69.662472  &   iras05278  & 980 & 9  & 1.20 ($K$) \\
IRAS 00554-7351, [GB98] S16            &  14.266458  & -73.587389  &   iras00554  & 720  & 8 & 0.72 ($K$) \\  
MSX LMC 1298, LMC134.1 40407           &  74.133958  & -68.880833  &  msxlmc1298  & 683 & pp & 1.12 ($I$) \\
MSX LMC  775, LMC166.2 19640, MACHO 8.8541.68 &  83.234083  & -68.213528  &   msxlmc775  & 2067, 2170 & pp, pp & 1.84 ($I$), 0.62 ($R$) \\
\hline
\end{tabular}

References: pp= present paper ; 1= Whitelock et al. (2003); 2= Groenewegen (2004); 3= Nishida et al. (2000); 
4= Raimondo et al. (2005); 5= Reid et al. (1988); 6= Sloan et al. (2006); 7= Wood (1998); 
8= Dodion (2003); 9= Wood et al. (2008, in prep.); 10= Groenewegen et al. (2007), 11= Hughes (1989);
12= Whitelock et al. (1994), 13= Wood et al. (1992); 14= Lloyd Evans (1985); 
15= Wood et al. (1983, periods actually from Payne-Gaposhkin, 1971);
16= Cioni et al. (2003); 17= Sloan et al. (2008); 18= Fraser et al. (2008); 
19= Pojmanski (2002, ASAS data in general)

\label{Tab-Csample}
\end{table*}

\begin{table*}

  \caption{The M-star sample: identifiers and pulsation periods for the first few entries.}
  \begin{tabular}{lrrlrrr}
  \hline
Names &  R.A.      & Declination & Identifier & Period & Ref. & Ampl. (Filter) \\
\hline

MSX LMC 1677, IRAS 06013-6505        &  90.365833  & -65.089750  &  msxlmc1677  & 348, 340 & pp, 19 & 2.1 ($I$)  \\
HD 271832, MSX LMC 1687, IRAS 06045-6722 &  91.106208  & -67.388444  &    hd271832  & 514, 52.7 & pp, 19 & 0.16 ($V$) \\
MSX LMC 1686                         &  91.699125  & -66.803472  &  msxlmc1686  & & &       \\
RS Men, IRAS 05169-7350, MSX LMC 412 &  78.921917  & -73.787139  &       rsmen  & 304 & 12 & 0.53 ($K$)  \\ 
WOH G  17, MSX LMC 1150              &  69.848708  & -73.184111  &      wohg17  & & &       \\
HD 269788, MSX LMC  778              &  83.723625  & -68.777639  &    hd269788  & 15.9 & pp & 0.007 ($I$) \\ 
\hline
\end{tabular}

\label{Tab-Msample}
\end{table*}

For all stars additional broad-band photometry ranging from the
optical to the mid-IR was collected from the literature,
primarily using
VizieR\footnote{http://vizier.u-strasbg.fr/viz-bin/VizieR} and the
NASA/IPAC Infrared Science
Archive\footnote{http://irsa.ipac.caltech.edu/}, using the coordinates
listed in Tables~\ref{Tab-Csample} and ~\ref{Tab-Msample}.  
In particular, we considered
\begin{itemize}

\item In the optical:
$UBVI$ data from Zaritsky et al.\ (2002, 2004) for the Magellanic Clouds (MCs).
$UBVR$ data from Massey (2002) for the MCs,
$BVRI$ data from Oestreicher et al.\ (1997) for RSG in the LMC,
OGLE $BVI$ data from Udalski et al.\ (1998),
$VRI$ data from Wood, Bessell \& Fox\ (1983, hereafter WBF).

\item In the near-infrared:
DENIS $IJK$ data from Cioni et al.\ (2000) and the third data release 
(The DENIS consortium 2005),
the all-sky $JHK$ release of 2MASS (Skrutskie et al.\ 2006), and the 
extended mission 6x long-exposure release,
$JHK$ data from the IRSF survey (Kato et al.\ 2007),
SAAO $JHKL$ data from Whitelock et al.\ (1989, 2003),
and CASPIR $JHKL$ data specifically taken for the IRS observations 
(Sloan et al.\ 2006, 2008, Groenewegen et al. 2007), and from Wood 
et al.\ (1992), Wood (1998).

\item In the mid-IR:
{\it IRAS} data from the Point Source Catalog, and the Faint Source 
Catalog (Moshir et al.\ 1989),
IRAC 3.6,4.5,5.8,8.0 and MIPS 24~\mum\ data from the SAGE catalog
(Meixner et al.\ 2006, first epoch data) and S$^3$MC catalog 
(Bolatto et al.\ 2007).

\end{itemize}

The literature considered is not exhaustive but does include all
recent survey data available in the near- and mid-IR, where 
these stars emit most of their energy.

\section{The model}
\label{themodel}

The models are based on the dust radiative transfer (RT) algorithm 
of Groenewegen (1993; also see Groenewegen 1995), which was 
developed to handle non-$r^{-2}$ density distributions in 
spherically symmetric dust shells.  The algorithm simultaneously 
solves the radiative transfer equation and the thermal balance 
equation for the dust.

The models for C stars begin with the stellar atmosphere models 
by Loidl et al.\ (2001; available for $T_{\rm eff}$ = 2650, 2800, 
3000, 3200, 3600 K), 
while for M stars the M0--M10 model atmospheres of Fluks et al.\ (1994)
are used.  These range from 3850 to 2500 K.  Strictly speaking,
the models are valid for giants at solar metallicities, but we have
applied them to Magellanic supergiants and AGB stars.
MARCS models\footnote{http://marcs.astro.uu.se/} are not yet available 
for abundances typical of AGB stars (i.e.\ with non-solar C/O ratios).

Our models assume that the dust around C stars is a combination of
amorphous carbon (AMC) and silicon carbide (SiC), with optical 
constants from, respectively, Rouleau \& Martin (1991; the AC1 
species), $\alpha$-SiC from P\'{e}gouri\'{e} (1988), and 
$\beta$-SiC from Borghesi et al.\ (1985), taking into account 
the matrix correction factors (see footnote in Groenewegen 1995).
These choices are based on the practical fact that these two types 
of dust fit actual observations.  We are aware of the discussion by, 
e.g., Pitman et al.\ (2008) about inadequacies in the derivation of 
optical constants in the literature, and that $\beta$-SiC is probably 
the primary carrier of the 11.3~\mum\ feature in C stars, rather 
than $\alpha$-SiC, in agreement with meteoritic data (see the 
discussion by Speck et al.\ 2009).  However, SiC features are known 
that clearly peak shortward of 11.3~\mum, and these are better
fitted with the constants of Borghesi et al.  Section~\ref{sec-c} 
discusses this point further.

Speck et al. also proposed that graphite rather then amorphous
  carbon dominates the dust, at least in the C-stars with
  extreme mass-loss rates they considered. Only one set of optical
  constants seems to have been published for graphite, those by Draine
  \& Lee (1984), as used by e.g. Volk et al. (1992), and Speck et al.
  Calculating the absorption coefficients for spherical grains using
  these optical constants results in a broad shoulder $\sim$~40 $\mu$m,  
  as is evident already in Fig.~4b in Draine \& Lee. This is not
  observed in C-stars, and is the reason why e.g. Martin \& Rogers
  (1987) already dismissed graphite in favour of AMC. Independently,
  graphite is expected to form at temperatures at high as 1800 K. As
  we will show below, when the condensation temperature is left as a
  free parameter it will typically be of order 1000 K, which is
  consistent with the condensation temperature of AMC.

Many C stars show a broad feature around 30~\mum\ which is believed
to arise from MgS dust (Goebel \& Moseley 1985).  Hony et al.\ (2002)
showed that the absorption coefficients depend strongly on the shape 
of the grains and that if MgS is the carrier, a continuous distribution
of ellipsoids (CDE) is required.
In addition they find that the temperature of the MgS grains is very
low, 100--400 K typically, and is thus unrelated to the warm dust
close to the star which is primarily responsible for the infrared
excess that we are modelling.  Consequently, we exclude this wavelength
region and MgS from our initial models.  
IRS observations confirm that MgS is associated with cool dust
temperatures, leading Zijlstra et al. (2006) to argue that the MgS
condenses onto existing grain surfaces.  Leisenring et al. (2008)
noted that the apparent SiC strength descreases as MgS increases, and
they suggested that the MgS coating hides the SiC emission feature.
In this scenario, our models would give only a lower limit to the
fraction of the SiC in the dust.  While we do not model the MgS, the
reader should keep in mind that its presence may mask spectral
features at shorter wavelengths.  
Section~\ref{secMGS} discusses the 30~\mum\ feature further.

For the M stars, several types of ``astronomical silicates'' are
available, as well as combinations of optical constants taken from
laboratory data.  For ``astronomical silicates'' we used the
absorption coefficients of Volk \& Kwok (1988, hereafter VK,
scaled down by a factor of 5 to agree with most other silicates), 
Draine \& Lee (1984, hereafter DL), 
``warm'' silicates from Suh (1999), 
Ossenkopf et al.\ (1992, hereafter OHM), 
David \& Pegourie (1995, hereafter DP), and combinations of DP with 
aluminium oxide (AlOx; amorphous porous Al$_2$O$_3$), with optical 
constants from Begemann et al.\ (1997), as used to model the SEDs of
M stars in the Galactic Bulge (Blommaert et al.\ 2006).

We constructed ``laboratory silicates'' using mixtures of olivine 
(Mg$_{0.8}$Fe$_{1.2}$ SiO$_4$, from Dorschner et al.\ 1995) and
AlOx and metallic iron (Ordal et al.\ 1988).  Harwit et al.\ (2001) 
and Kemper et al.\ (2002) have advocated the use of metallic iron
to increase the opacity in the near-IR region, and the present
models confirm this need.  We used discrete combinations with 3, 4 
or 5\% Fe, and 0, 10, 20, 30 or 40\% AlOx, with olivine accounting 
for the remainder.

Other dust components have been identified in the spectra of M-stars,
e.g.\ spinel or corundum (Posch et al.\ 1999; Sloan et al.\ 1996), 
but these are only minor components and remain controversial (e.g., 
Sloan et al.\ 2003, Heras \& Hony 2005, Depew et al.\ 2006). 
The aim of the present paper is to {\it globally} fit the SED as
opposed to fitting the details of the IRS spectra of the stars.

For the dust mixtures in both C and M stars, the extinction
coefficients have been calculated in the small-particle limit, and
($Q_{\lambda}/a$) is calculated from adding in proportion the
extinction coefficients of the individual species.  In other words, 
the grains are treated separately, but all have the same temperature. 
The standard model does not consider core-mantle grains, and the
composition of the grains is independent of distance to the star. 
See section~\ref{sec-m} for further discussion.

Distances of 50 kpc to the LMC and 61 kpc to the SMC are adopted.
The models have been corrected for a typical $A_{\rm V}$ = 0.15 for 
all stars.  The exact value is of little importance as this 
corresponds to \less 0.02 mag of reddening in the near-IR.
For all stars an expansion velocity, v$_{\rm exp}$, of 10~\ks\ and a 
dust-to-gas ratio of 0.005 have been adopted.

We fitted the models to the SEDs in the following manner.
The RT model was included as a subroutine in a
minimisation code using the the {\sc mrqmin} routine (using the
Levenberg-Marquardt method from Press et al.\ 1992).  Parameters that
were fitted in the minimisation process include the mass-loss rate 
(\mdot), luminosity, and the dust temperature at the inner radius 
($T_{\rm c}$), although we sometimes fixed \mdot\ and/or $T_{\rm c}$.
The output of a model is an SED, and this is folded with the relevant 
filter response curves to produce magnitudes to compare to the
observations (see Groenewegen 2006).

\begin{figure} 

\begin{minipage}{0.49\textwidth}
\resizebox{\hsize}{!}{\includegraphics{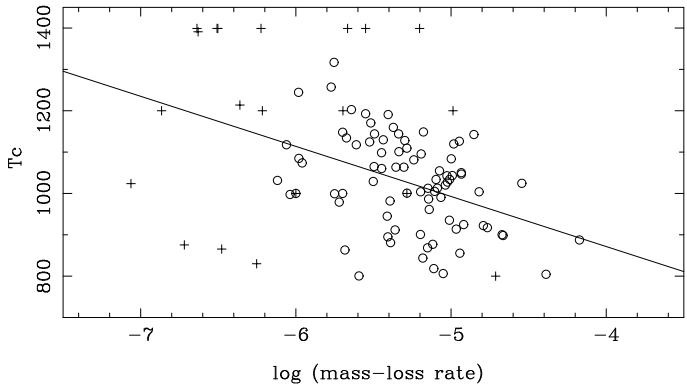}}
\end{minipage}


\caption[]{ 
Dust temperature at the inner radius versus mass-loss rate for C stars.
Plus-signs indicate stars excluded from the fit (the solid line, Eq.~\ref{Eq-ML}).
} 
\label{Fig-Tc-ML} 
\end{figure}

Typically, for a fixed effective temperature the model was minimised
for the different grain types, for both fitted and fixed $T_{\rm c}$.
For the grain type with the lowest $\chi^2$ error, the model was run 
for different model atmospheres.  This is reasonable because the IRS 
spectrum largely determines the best-fitting model atmosphere, while the
effective temperature is largely determined from the optical and near-IR
broad-band photometry.  When fixed, $T_{\rm c}$ was set at temperatures
typically expected, i.e.\ 900, 1000, 1100, 1200 K.  For C stars it has
been suggested (e.g.\ Groenewegen 1995) that $T_{\rm c}$ decreases with 
optical depth.  This effect has been investigated by first allowing 
$T_{\rm c}$ to vary.  Excluding mass-loss rates below 6. 10$^{-7}$ 
\msolyr\ where $T_{\rm c}$ and \mdot\ are not simultaneously fitted 
well, and stars where the fit converged to temperatures above 1400 K 
or below 800 K, a linear relation gives
\begin{equation}
 T_{\rm c} = (-121 \pm 29) \log \dot{M} + (386 \pm 156)
\label{Eq-ML}
\end{equation}
based on 82 stars, with a mean of 1030 K, see Fig.~\ref{Fig-Tc-ML}.
In subsequent model runs, $T_{\rm c}$ was then also fixed to the 
value given by Eq.~\ref{Eq-ML}.
For M stars such a dependence has never been suggested and has not 
been investigated.

For some C-stars it also turned out that lowering the outer radius of
the dust shell markedly improved the fit.  Groenewegen et al.\ (2007)
discussed two cases previously, and Sloan et al.\ (2009) found this
necessary for a C star in the Sculptor Dwarf.  The default outer radius 
(as a multiple of the inner radius) is typically of order 10~000 and is
determined by considering a dust temperature at the outer radius of 20
K.  For stars where a smaller outer radius seemed necessary, new models
were run decreasing the outer radius by a factor of 3 each time until
the value of $\chi^2$ rose again.  For none of the M stars a smaller
outer radius was necessary, but for a non-negligible fraction of the
C stars (20\%), it was.  Although not explicitly tested, the same net 
effect (less flux at longer wavelengths) can be achieved by a steeper 
density law than the $r^{-2}$ assumed in the RT model.

The quality of the fit is based on a $\chi^2$ analysis
\begin{equation} 
  \chi^2= \sum_{i=1}^{i=n} (m_{\rm obs}(i) - m_{\rm pred}(i))^2 / (\sigma_{\rm m_{obs}(i)})^2,
\label{eq-chi}
\end{equation} 
with $m$ the observed or predicted magnitude (for the broadband
photometry) or flux (for the spectrum) with error bar $\sigma_{\rm m_{obs}}$ 
and $n$ is the total number of measurements.  Also the reduced $\chi^2$ is 
defined:
\begin{equation} 
  \chi_{\rm r}^2 = \frac{\chi^2}{(n-p)}, 
\end{equation} 
with $p$ the number of free parameters, and the quantity
\begin{equation} 
  {\rm BIC} = \chi^2 + (p + 1)\; \ln (n).
\end{equation} 
This is based on the Bayesian information criterion (Schwarz 1978) and
measures whether an increase in the number of free
parameters and the resulting lower $\chi^2$ is actually significant. 
In this way it is possible, for example, to compare models with
fitted and fixed $T_{\rm c}$, by comparing the values of BIC.

In Eq.~\ref{eq-chi} broadband data and spectral data are included in
one formalism.  However it is not evident how to weigh the typically
10--40 available photometric data points with the $\sim$350 spectral data points.
For a few stars, where the fit to the SED alone already provided a
good match to the IRS spectrum both in absolute flux and spectral
shape the following experiment was done.  The IRS spectrum covers the 5
to 38~\mum\ region, and broad-band photometry available in this region
(IRAC 5.8 and 8.0~\mum, MSX A, MIPS 24~\mum, and IRAS 12 and 25~\mum\
bands) has a typical resolution of order 4--6.  At such a resolution,
10--13 such filters fit within the 5--37~\mum\ region covered by the  IRS.
%
In the experiment we added, effectively duplicated, 10--13 of the
available IRAC, MSX, MIPS, and IRAS, data points and calculated the 
resulting reduced $\chi^2$.  Then the original photometric dataset was 
fitted including the IRS spectrum, and the uncertainties of the spectral 
data were scaled so as to produce the same reduced $\chi^2$.  We
examined four stars in this manner, found the scaling factors to be
in the range 3--8, and concluded that the uncertainties of the IRS
spectral data should be scaled by a factor of 5.

We masked those portions of the IRS spectra with poor S/N or those
affected by background subtraction problems and did not include them
in the minimisation procedure.  In addition, regions where strong 
molecular features dominate that are not included in the simple 
C-star model atmospheres are also excluded for the C stars, i.e.\ the 
regions
4.3--6.2~\mum\ (CO + C$_3$, see e.g.\ J\o rgensen et al.\ 2000), 
6.6--8.5~\mum\ and 
13.5--13.9~\mum\ (C$_2$H$_2$ + HCN band, see e.g.\ Matsuura et al.\ 2006), and
22.0--38.0~\mum\ (the region of the MgS dust feature, which is not considered).

\section{Results} 

\begin{figure*}
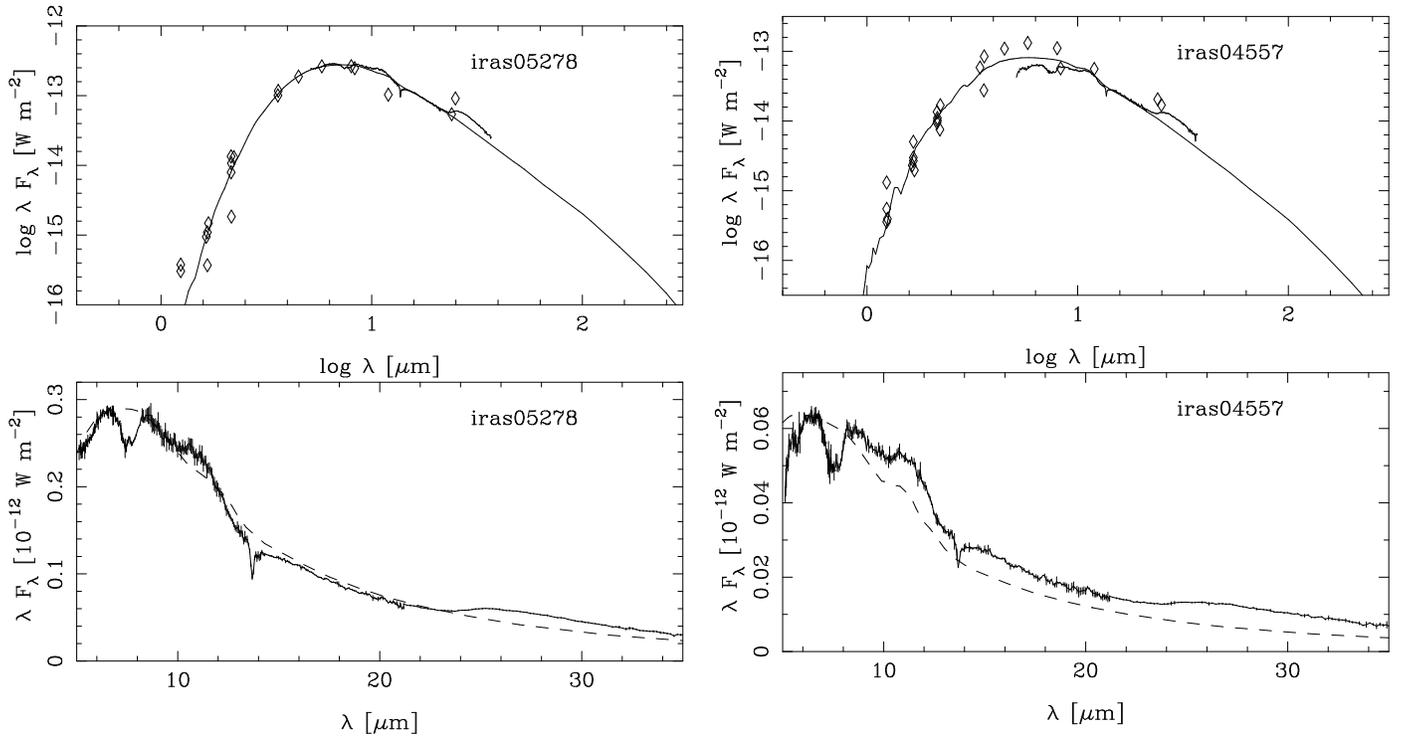
 

\begin{minipage}{0.49\textwidth}
\resizebox{\hsize}{!}{\includegraphics[angle=-90]{rt_iras05278.ps}}
\end{minipage}
\hfill
\begin{minipage}{0.49\textwidth}
\resizebox{\hsize}{!}{\includegraphics[angle=-90]{rt_iras04557.ps}}
\end{minipage}

\caption[]{ 
Fits to the SEDs and IRS spectra of C-stars. 
This figure is available in the electronic edition.
} 
\label{Fig-Cstars} 
\end{figure*}

\begin{figure*}
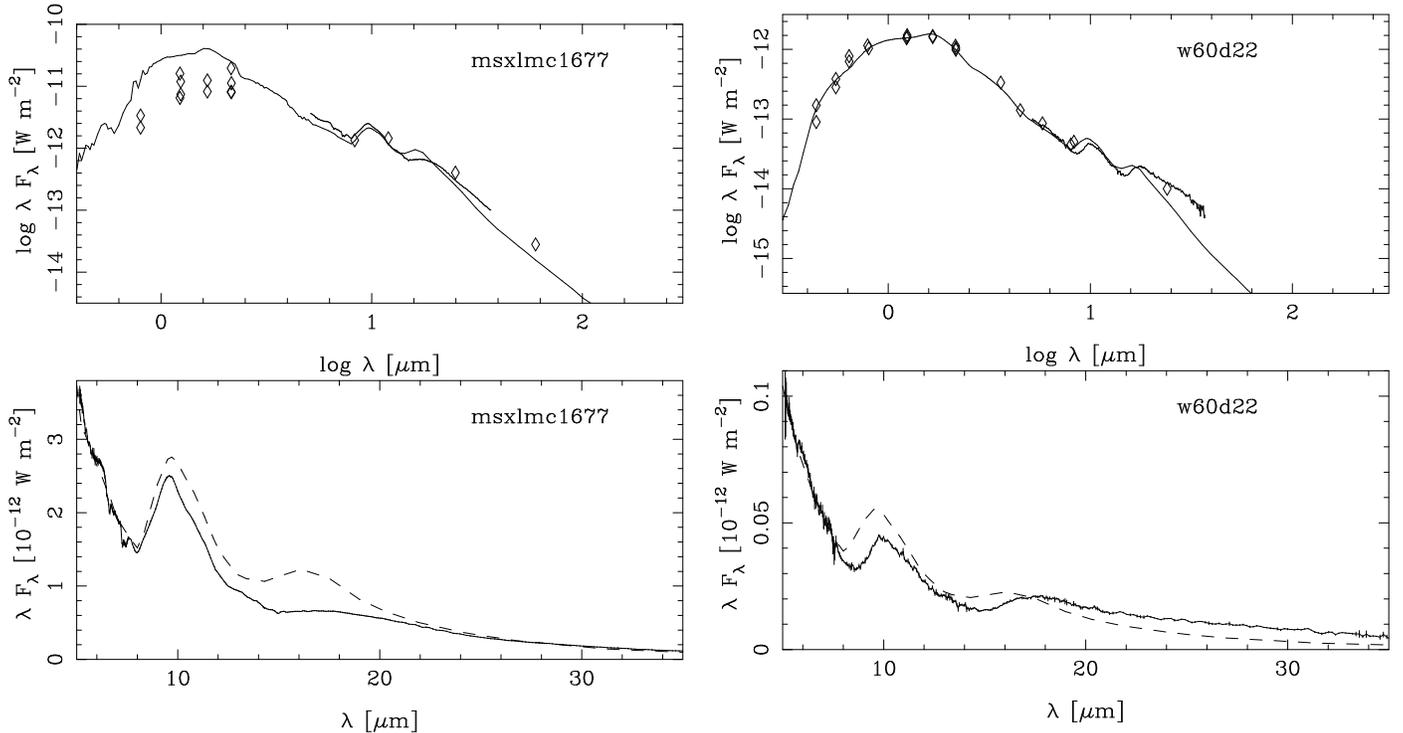
 

\begin{minipage}{0.49\textwidth}
\resizebox{\hsize}{!}{\includegraphics[angle=-90]{rt_msxlmc1677Lab.ps}}
\end{minipage}
\hfill
\begin{minipage}{0.49\textwidth}
\resizebox{\hsize}{!}{\includegraphics[angle=-90]{rt_w60d22Lab.ps}}
\end{minipage}

\caption[]{ 
Fits to the SEDs and IRS spectra of M-stars using ``laboratory silicates''.
This figure is available in the electronic edition.
} 
\label{Fig-MLab} 
\end{figure*}

\begin{figure*}
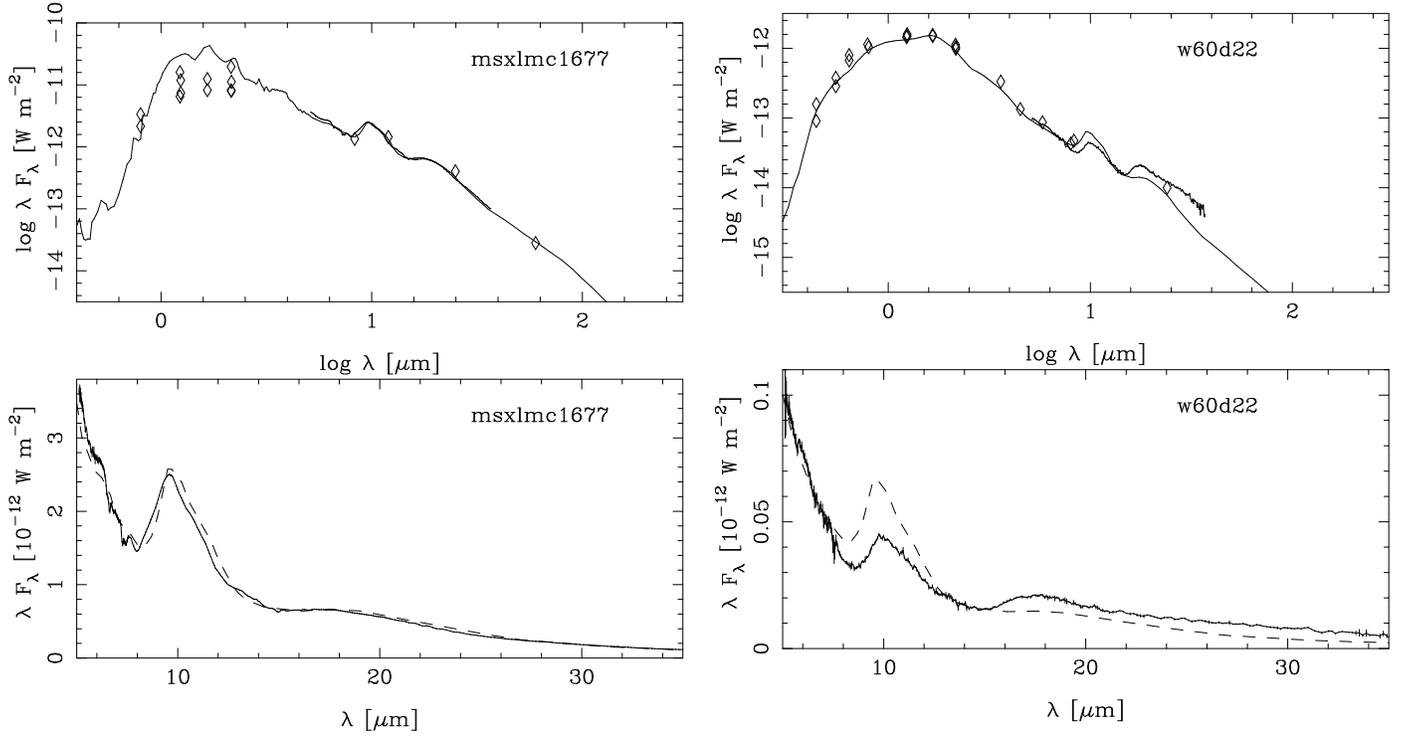
 

\begin{minipage}{0.49\textwidth}
\resizebox{\hsize}{!}{\includegraphics[angle=-90]{rt_msxlmc1677Ast.ps}}
\end{minipage}
\hfill
\begin{minipage}{0.49\textwidth}
\resizebox{\hsize}{!}{\includegraphics[angle=-90]{rt_w60d22Ast.ps}}
\end{minipage}

\caption[]{ 
Fits to the SEDs and IRS spectra of M-stars using ``astronomical silicates''.
This figure is available in the electronic edition.
} 
\label{Fig-MAst} 
\end{figure*}

Tables~\ref{Tab-Cstar} and \ref{Tab-Mstar} lists the parameters of 
the models which best fit the observed data.  (Only the first entries are
shown for guidance.  Both tables are fully available at the CDS).  For the
M stars, the first line lists the best overall fit, while the second 
lists the best fit with an ``astronomical silicate'' if the overall best 
fit is for a ``laboratory silicate'', and vice versa.
Listed are the identifier, effective temperature or spectral type,
dust type, luminosity, mass-loss rate, whether \mdot\ was fitted (1)
or fixed (0), $T_{\rm c}$, whether $T_{\rm c}$ was fitted or fixed, 
outer radius (in units of inner radius), dust optical depth in the $V$-band,
optical depth at 11.3 $\mu$m (C-stars), or the peak of the silicate
feature (This depends on the type of silicate used, but is near 10 $\mu$m), 

Figures~\ref{Fig-Cstars}, ~\ref{Fig-MLab} and ~\ref{Fig-MAst} in the 
electronic edition show the best fits, for the M stars again for both
``astronomical silicates'' and ``laboratory silicates''.
The top panel shows the observed SED and IRS spectrum and the 
fitted model on an absolute scale, while the bottom panel shows the IRS 
spectrum, scaled to a quasi-continuum point based on the 
average flux in the 6.4--6.5~\mum\ region.

The fitting routine also provides uncertainties for the mass-loss 
rate, dust temperature at the inner radius, and luminosity.  These 
are typically small, of order 1\%.  The true errors are larger, and 
can be estimated from a comparison of model runs with different 
parameters.  They are typically 10\% in luminosity, 25\% in 
mass-loss rate and 50 K in $T_{\rm c}$.


\begin{table*} 
\caption{Fit results of the C-star sample for the first few entries} 
\begin{tabular}{rrrrrrrrrrrrr} \hline \hline 
Identifier  & $T_{\rm eff}$ &  grain &  $L$    & \mdot     &  fit? & $T_{\rm c}$ & fit? & $R_{\rm out}$ & $\tau_{0.5}$ & $\tau_{11.3}$ &  \\
            &     (K)      &  type  & (\lsol) & (\msolyr) &       &    (K)     &      &             &              &              &  \\
\hline 
  iras04496 & 2800 &   sic0 &   35690 & 5.18e-06 & 1 & 1000 & 0 & 10000.0 &  4.646 &  0.145 &    \\
  iras05278 & 3600 &   sic0 &   30617 & 2.85e-05 & 1 & 1055 & 0 &  6000.0 & 22.645 &  0.704 &    \\
  iras00554 & 3600 &   sic4 &   27127 & 1.42e-05 & 1 & 1204 & 0 & 10000.0 & 16.465 &  0.752 &   \\
 msxlmc1298 & 3600 &   sic4 &   24506 & 1.15e-05 & 1 & 1142 & 0 &    41.2 & 12.977 &  0.593 &    \\
  msxlmc775 & 3200 &  bsic2 &   20776 & 5.59e-06 & 1 & 1082 & 1 &     4.6 &  7.170 &  0.284 &   \\
\hline 
\end{tabular}

sic refers to $\alpha$-SiC from P\'{e}gouri\'{e} (1988), bsic to $\beta$-SiC from Borghesi et al. (1985). 
The number refers to the percentage of SiC. The rest is amorphous carbon from Rouleau \& Martin (1991) 
(also see Sect.~\ref{themodel}).

\label{Tab-Cstar}
\end{table*}



\begin{table*} 
\caption{Fit results of the M-star sample for the first few entries} 
\begin{tabular}{rrrrrrrrrrrrr} \hline \hline 
Name        & Spectral &  grain &  $L$    & \mdot     &  fit? & $T_{\rm c}$ & fit? & $R_{\rm out}$ & $\tau_{0.5}$ & $\tau_{10}$ &  \\  
            & type (K) &  type  & (\lsol) & (\msolyr) &       &    (K)     &      &              &              &            &  \\
\hline 
 msxlmc1677 &  m8 &              VK           & 2773703 & 2.63e-06 & 1 &  802 & 1 & 10000.0 &  0.249 &  0.056 &    \\ 
 msxlmc1677 &  m5 &     sil75alox20fe5        & 3194712 & 2.06e-06 & 1 &  918 & 1 & 10000.0 &  0.106 &  0.034 &   \\ 
   hd271832 &  m6 &              VK           & 1787262 & 2.00e-08 & 0 &  900 & 0 & 10000.0 &  0.003 &  0.001 &    \\ 
   hd271832 &  m6 &     sil65alox30fe5        & 1791278 & 1.60e-08 & 0 &  900 & 0 & 10000.0 &  0.001 &  0.000 &    \\ 
 msxlmc1686 & m10 &           DP 0.8 alox 0.2 & 1215022 & 9.42e-07 & 1 & 1000 & 0 & 10000.0 &  0.134 &  0.032 &   \\ 
 msxlmc1686 & m10 &     sil55alox40fe5        & 1235338 & 6.02e-07 & 1 &  900 & 0 & 10000.0 &  0.056 &  0.018 &    \\ 
      rsmen &  m9 &              VK           &  724206 & 5.23e-07 & 1 & 1000 & 0 & 10000.0 &  0.156 &  0.035 &   \\ 
      rsmen &  m9 &     sil65alox30fe5        &  747279 & 5.74e-07 & 1 &  900 & 0 & 10000.0 &  0.073 &  0.024 &    \\ 
\hline 
\end{tabular} 

Grain type refer to: VK = Volk \& Kwok (1988), DL = Draine \& Lee (1984),  Suh = Suh (1999), OHM = Ossenkopf et al. (1992), 
DP = David \& Pegourie (1995), and combinations of DP with Aluminium Oxide (AlOx).
The types marked silXaloxYfeZ, refer to X-percent olivine, Y-percent AlOx, and Z-percent metallic iron.
(see Sect.~\ref{themodel} for details).

\label{Tab-Mstar}
\end{table*}

\section{Discussion} 

\subsection{Foreground objects and the division between AGB stars 
and supergiants}
\label{sec-fg}

The brightest M stars in the sample have luminosities in excess of one
million solar luminosities if they are indeed in the MCs.  The present
section tries to identify likely foreground (FG) objects.  In addition
the separation between RSGs and (oxygen-rich) AGB stars is of interest.
WBF separated RSGs and AGB Stars based on (1) the location of a source
in a diagram plotting $M_{\rm bol}$ versus period, and (2) the
amplitude of pulsation, with AGB stars showing larger amplitudes.
Smith et al.\ (1995) used a similar diagram, and they also considering
the presence or absence in the atmosphere of lithium, which indicates
that the star has experienced hot bottom burning (HBB) and is thus at
the upper end of the mass range for stars on the AGB.

\begin{figure}

\begin{minipage}{0.49\textwidth}
\resizebox{\hsize}{!}{\includegraphics{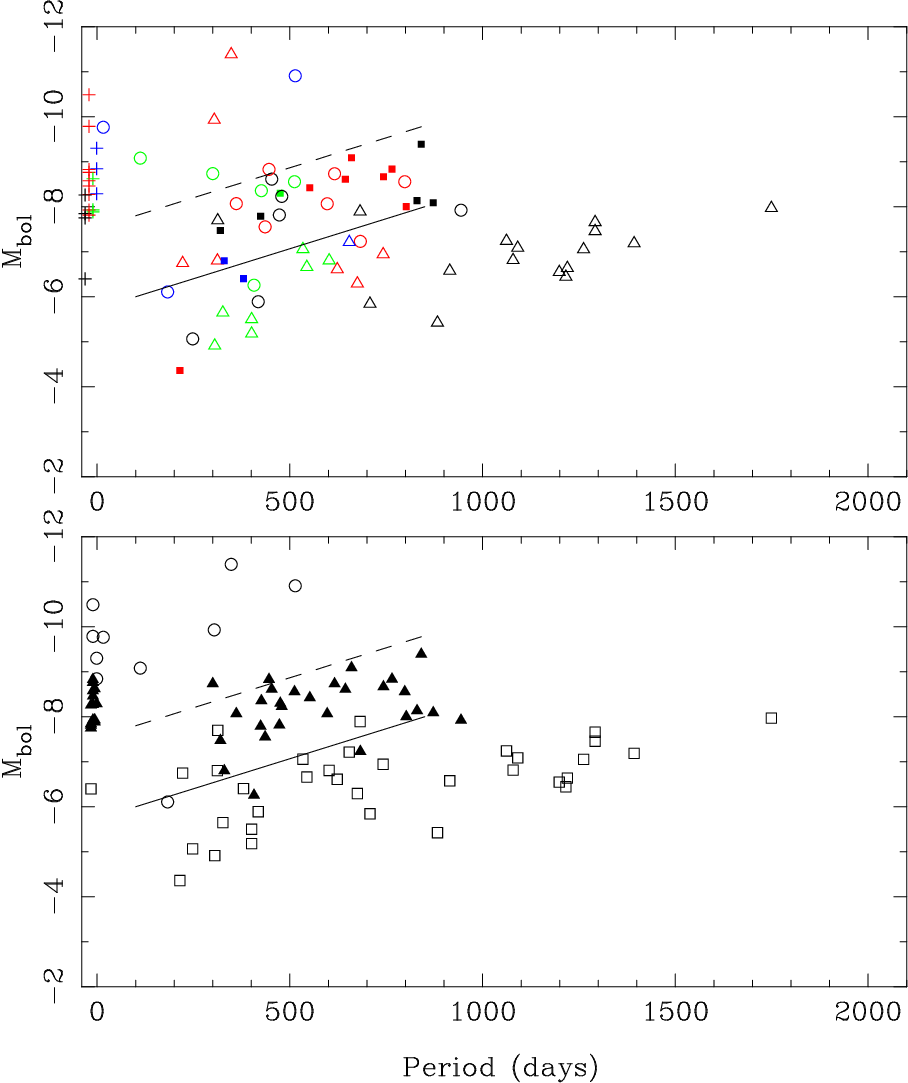}}
\end{minipage}

\caption[]{ 
Top panel. Bolometric magnitude versus pulsation period for the M stars.
Stars without period are plotted as plus signs at negative periods.
The full line indicates the lower luminosity limit for RSGs by WBF, and 
the dashed line is 1.8 mag brighter.
Top panel.
$I$-band semi-amplitudes larger than 0.45 magnitudes are denoted by 
triangles.
Amplitudes between 0.2 and 0.45 by filled squares.
Amplitudes smaller then 0.2  by circles.
In the electronic edition, [3.6]$-$[8.0] colours redder than 1.2 are 
shown in black, between 0.5 and 1.2 in red, between 0.15 and 0.5 in 
green, and less than 0.15 in blue. 
Bottom panel.  As top panel, but the objects are identified as
foreground objects (open circles),  RSG (filled triangles), and 
AGB stars (open squares) based on the discussion in 
Sect.~\ref{sec-fg}. 
} 
\label{Fig-FG} 
\end{figure}

Figure~\ref{Fig-FG} shows this diagram, using different symbols for
different $I$-band amplitudes (see caption).  Stars without (known) 
period are plotted as crosses at negative period.
Amplitudes in $B,V,R,$ and $K$ have been multiplied by, respectively, 0.14,
0.18, 0.66, and 2.3 to estimate the amplitude in $I$. These ratios are
based on values actually determined for the stars in our sample.
Colour is another useful criterion, as it indicates the presence of 
circumstellar material.  In the electronic edition, [3.6[$-$[8.0] 
colours are coded by different colours (see caption).

A check of the literature indicates two obvious FG objects. 
RS Men has a radial velocity of 140 \ks, far lower then the
250--300 \ks\ typical of LMC objects. 
HD 269788 is listed as having a significant proper motion, and has 
a spectral type of K4 III.
None of the other stars have spectral types, radial velocities, 
parallaxes or proper motion that would suggest a foreground nature.

HD 271832 is another likely FG object. The object is plotted at its
period (514 days) which shows the largest amplitude (0.16 in $V$).
This may not be its pulsation period but possibly related to the LSP
(long secondary period) phenomenon observed in AGB stars (see
e.g.\ Soszy\'nski 2007 and references therein).  The object also shows
a period of 52.7 days with an amplitude of 0.06 mag.  This period and
amplitude suggest that the object is a probable early-type giant in
the FG.

Early-type FG stars are expected to have essentially no mass-loss rate
(hereafter MLR) and higher temperatures than AGB stars.  HD 269788 and
HD 271832 have a fitted optical depth at 10~\mum\ of $\le$ 0.001, and
a [3.6]$-$[8.0] colour $\le$ 0.09.  Three other stars fulfill both
criteria, and are also classified as FG:  W60 D29, MSX LMC 1212 and HV
11366. The last object has a period of 183 days (derived from MACHO
data), but a small amplitude.  WBF reported a spectral type of MS and
a period of 366 days for HV 11366, and based on this period Sloan et
al.\ (2008) assumed it was a member of the SMC.  The period quoted by
WBF comes from Payne-Gaposhkin (1971). Either the shorter period was
missed,  
or the star has switched from fundamental to overtone mode.

Buchanan et al.\ (2006) identified RS Men as a FG object because
it is a Mira variable and its luminosity is not consistent 
with the Mira period-luminosity relation (Feast et al.\ 1989) if
it were at the distance of the LMC, corroborating the FG nature
suggested by its radial velocity.  Buchanan et al.\ also used
the $PL$ relation to identify MSX LMC 1677 and MSX LMC 1686 as FG 
Mira variables.  If MSX LMC 1686 were in the LMC, its luminosity 
would exceed that of a very bright RSG ($L \approx$ 1.2 million
\lsol).  Similarly, WOH G 17 would have have a luminosity of
600~000 \lsol, and thus we suspect that it is also FG.
The circumstellar reddening and the [3.6]$-$[8.0] colours of
MSX LMC 1686 and WOH G 17 are consistent with this suspicion.

MSX LMC 946 is a SR with a period of 112 days.  If it were in 
the LMC, it would also be too bright (with $L \approx$ 300~000 \lsol).
We believe it also is a FG star.

Both WBF and Smith et al.\ (1995) consider the maximum luminosity of
an AGB star to be $M_{\rm bol} = -7.1$.  This limit is based on the
maximum possible core mass of 1.4 \msol\ before core He ignition and
the core-mass luminosity relation of Paczy\'nski (1970) or Wood \&
Zarro (1981).  Other limits have also been suggested.  Wagenhuber \&
Groenewegen (1998) give a limit of $L$ = 16~400, 31~800 and 119~000
\lsol\ for $M_{\rm c}$= 0.8, 1.0 and 1.4 \msol\ or $M_{\rm bol} =
-8.0$, respectively (although none of the full evolutionary
calculations on which all of these core-mass luminosity relations are
based have actually evolved a star to the Chandrasekhar limit).
Poelarends et al.\ (2008) in their recent study of super-AGB stars
examined a 9~\msol\ star.  Their model had a core mass of 1.17~\msol\ 
after the second dredge-up at the start of the first thermal
pulse, and it reached $\log L$ = 5.07 ($L$= 118~000 \lsol, $M_{\rm bol} 
= -8.0$) at the twelfth pulse shortly after.
These limits are consistent with the bolometric magnitude of the 
longest-period variable in the sample, MSX SMC 055, which
has a period of 1749 days and an $I$-band amplitude of 0.81.
Among the stars in the sample, this object is 
the most likely candidate for a super-AGB star.
%
%

Stars (with or without period) brighter than $M_{\rm bol} = -8.0$ are
classified as RSG.  All of these turn out to have amplitudes lower than
that expected for Mira variability.  Therefore all stars with
amplitudes smaller than 0.45 and in between the solid and dashed line
in Fig.~\ref{Fig-FG} are classified as RSG, and all remaining stars
with amplitudes larger than 0.45 are classified as AGB stars.

Four stars with a known period have uncertain classifications.  MSX LMC 
1318 and HV 11223 are slightly fainter than the lower limit used by WBF, 
but have blue colours and small amplitudes and are tentatively 
classified as RSGs.  MSX SMC 134 has a small amplitude but reasonable 
red colours and is suspected to be an AGB star.  Similarly, WBP 77 has
an amplitude close to that of a Mira variable and is treated as an
AGB star.

Of the remaining stars without period, MSX LMC 807 is fainter than
$M_{\rm bol} = -6.4$, has a red colour, [3.6]$-$[8.0] $>4$, and is
classified as an AGB star.  The remainder are brighter than 
$M_{\rm bol} = -7.8$, have blue colours, [3.6]$-$[8.0] $<0.1$, and are
classified as RSG.

We have separated RSGs and AGB stars primarily to
study the mass-loss process in these two classes of objects (see 
Sect.~\ref{sec-ML}).  However, the conclusions drawn there are not
sensitive to the uncertainties in classification described here.
\bigskip

The brightest C star in the sample has a luminosity of 35~000 \lsol, 
or $M_{\rm bol}= -6.6$.  The faintest has 4~100 \lsol, or 
$M_{\rm bol} = -4.3$. The brightest C star in the SMC is the third
brightest overall.
The largest MLR is 7.1 10$^{-5}$ \msolyr, for the seventh most luminous C star.
Although generally speaking, a lower luminosity implies a smaller 
MLR, there are some exceptions.  The fourth smallest MLR in the 
sample, 1.8 10$^{-7}$ \msolyr, is for CV 78, with the thirteenth highest 
luminosity of 16~000 \lsol.
\bigskip

\begin{figure} 

\begin{minipage}{0.49\textwidth}
\resizebox{\hsize}{!}{\includegraphics{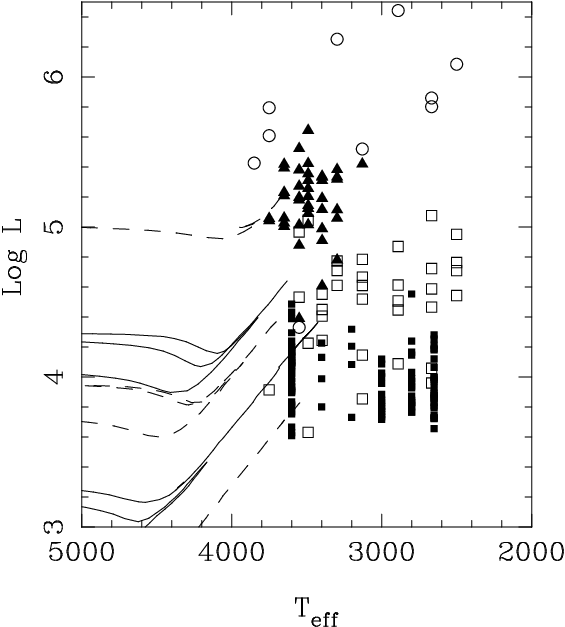}}
\end{minipage}

\caption[]{ 
The Hertzsprung-Russell diagram.
C-stars are plotted as filled squares, M-type AGB stars as open squares, 
RSGs as filled triangles, and foreground objects as open circles.
Lines indicate evolutionary tracks by Bertelli et al.\ (2008) for 
20, 10, 8, 5, and 3 \msol\ (top to bottom) for $Z$= 0.008.
} 
\label{Fig-LT} 
\end{figure}

Figure~\ref{Fig-LT} shows the Hertzsprung-Russell diagram with
evolutionary tracks of Bertelli et al.\ (2008), extended to unpublished
tracks of more massive stars (Girardi, priv. comm.) for $Z$= 0.008.  
For the intermediate-mass stars the evolution is stopped at the
beginning of the AGB, which is why the tracks do not pass through the
cloud of points marking the O- and C-rich AGB stars.  Most of the RSGs
are consistent with the 10--20~\msol\ tracks (cf. the discussion on
the revised spectral type effective temperature calibration in
Levesque et al. 2005, 2006).
The coolest RSG is MSX LMC 891 for which we assign a spectral type of M7 (3129 K).

\subsection{Colour-magnitude and colour-colour diagrams}

\begin{figure*}
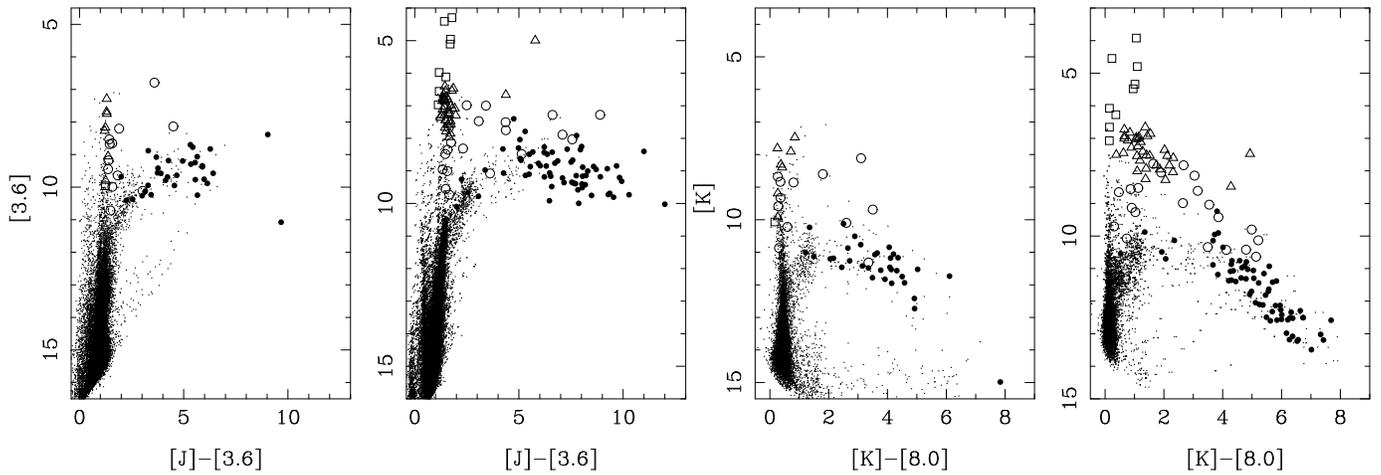


\begin{minipage}{0.49\textwidth}
\resizebox{\hsize}{!}{\includegraphics{sirtf_hrd5_PAPER.ps}}
\end{minipage}
\begin{minipage}{0.49\textwidth}
\resizebox{\hsize}{!}{\includegraphics{sirtf_hrd4_PAPER.ps}}
\end{minipage}

\caption[]{ 
Colour-magnitude diagrams from two combinations of colours for the 
SMC (left-hand panel) and LMC (right-hand panel).
Dots are stars picked randomly from the S$^3$MC and SAGE surveys.
M stars are plotted as squares (foreground objects), triangles 
(supergiants), and circles (AGB stars).
C stars are plotted as filled circles.
} 
\label{Fig-HRD} 
\end{figure*}

Figure~\ref{Fig-HRD} shows two colour-magnitude diagrams based on IRAC
and NIR colours.  The background of stars plotted as dots are taken 
from a random subset of S$^3$MC and SAGE data (and having errors in 
the magnitudes of less than 0.1 mag).  The photometry plotted
for the sample is the synthetic photometry calculated for the best fit
model, and not the observed magnitudes.

Matsuura et al.\ (2009), Vijh et al.\ (2009), and others  
have recently published colour-magnitude diagrams with IRAC and/or NIR 
colours showing regions where particular types of objects can be found
or overplotting stars with known spectral types.
Our results agree with these previous works. 


\begin{figure*}
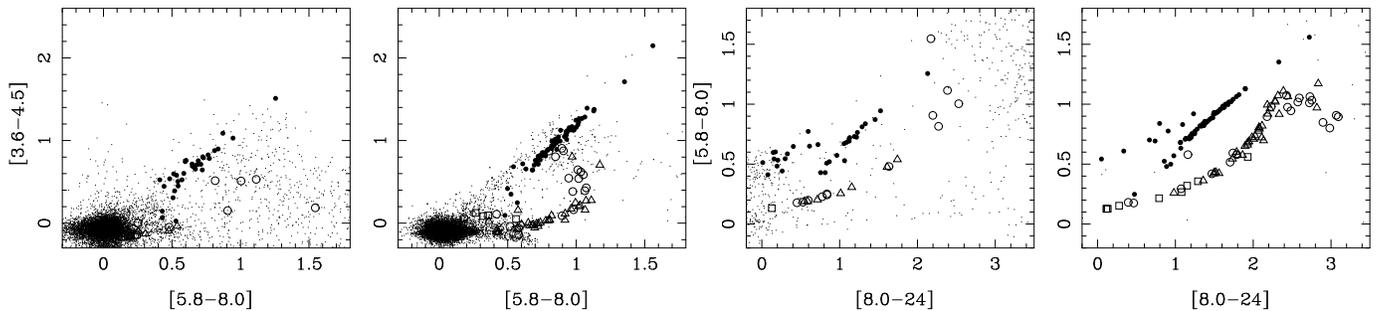


\begin{minipage}{0.49\textwidth}
\resizebox{\hsize}{!}{\includegraphics{sirtf_ccd3_PAPER.ps}}
\end{minipage}
\begin{minipage}{0.49\textwidth}
\resizebox{\hsize}{!}{\includegraphics{sirtf_ccd4_PAPER.ps}}
\end{minipage}

\caption[]{ 
Colour-colour diagrams for two combinations of colours for the SMC 
(left-hand panel) and LMC (right-hand panel).
Dots are from the S$^3$MC and SAGE survey.
M stars are plotted as squares (foreground objects), triangles 
(supergiants), and circles (AGB stars).  
C stars are plotted as filled circles.
} 
\label{Fig-CCD} 
\end{figure*}

Figure~\ref{Fig-CCD} shows two colour-colour diagrams, in particular
those that Kastner et al.\ (2008) indicate are effective in
distinguishing M from C stars.  We confirm this, in particular for 
the [5.8]$-$[8.0] vs.\ [8]$-$[24] diagram. 


\begin{figure*}
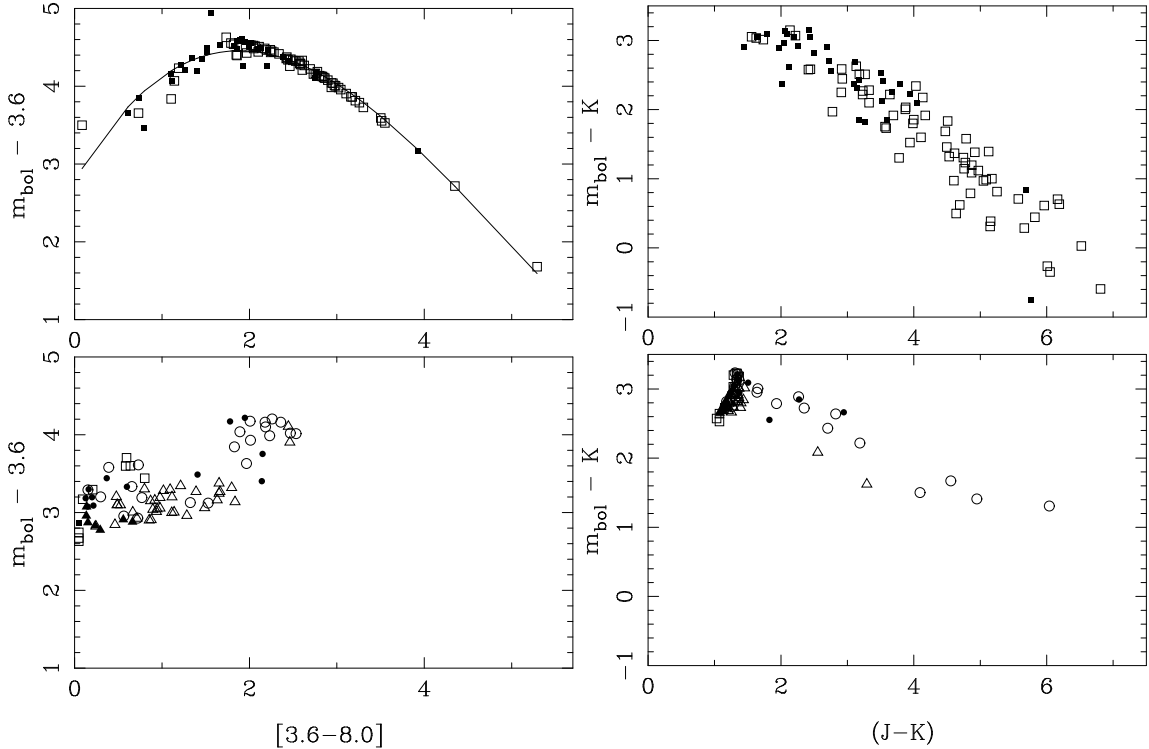


\begin{minipage}{0.41\textwidth}
\resizebox{\hsize}{!}{\includegraphics{BC_col-paper.ps}}
\end{minipage}
\begin{minipage}{0.41\textwidth}
\resizebox{\hsize}{!}{\includegraphics{BC_col1-paper.ps}}
\end{minipage}

\caption[]{ 
The bolometric correction at 3.6 \mum\ versus [3.6]$-$[8.0] colour 
(left) and $K$ versus $J-K$ colour (right) 
for C stars (top panel) and M stars (bottom panel).
SMC objects are plotted as filled squares in the top panel.
In the bottom panel, foreground objects are plotted as squares, 
supergiants as triangles and AGB stars as circles.
SMC objects are plotted with filled symbols.
The curved line in the top right panel is the fit to the data: 
$m_{\rm bol} = 2.78 + 1.926 \, ([3.6-8.0]) -0.6309  \, ([3.6-8.0])^2  +0.04239 \, ([3.6-8.0])^3$
} 
\label{Fig-BC} 
\end{figure*}

Figure~\ref{Fig-BC} shows the bolometric correction (BC) at 3.6
\mum\ versus [3.6]$-$[8.0] colour, and at $K$ versus $J-K$ colour for
the synthetic colours (also see WBF and Whitelock et al.\ 2003 for BCs
involving $K$ and other colours).
Relations like those presented here make it is possible to estimate
bolometric magnitudes with an estimated uncertainty of about 0.2 mag,
which might be sufficient for many applications.  Such an estimate
could even serve as a starting point for more detailed modelling.

\subsection{Dust in the M-stars}
\label{sec-m}

The primary aim of the present paper is to derive luminosities and
MLRs by fitting models to the broad-band data, as opposed to
fitting the IRS spectra in detail.  Nonetheless, some interesting 
remarks can and should be made.

We have considered two families of oxygen-rich dust.  The first is the
``astronomical silicates.''  These are derived empirically from observed 
spectroscopic data, generally galactic AGB stars.  The second family
is based on the optical constants measured in the laboratory, denoted
here as ``laboratory silicates''.  Figures~\ref{Fig-MLab} and ~\ref{Fig-MAst} 
show the best fitting models determined independently for the two families.

Of the 86 M stars, the MLR is fitted in 75 objects, and in 57 of
those cases, an astronomical silicate provides the best overall fit.  
In most of those cases (70\%), the best fit comes from the
``astronomical silicate'' produced by Volk \& Kwok (1988), which is
based on data from the Low-Resolution Spectrograph and photometry at
12, 25, 60, and 100~\mum\ from {\it IRAS} of oxygen-rich AGB stars.

An inspection of Figs.~\ref{Fig-MLab} and ~\ref{Fig-MAst} 
illustrates, however, that the best fitting models often give a
far from perfect match to the data.
In the astronomical silicates the 18~\mum\ feature is often 
too weak w.r.t.\ the 10~\mum\ feature, and broader than 
observed,  Additionally, the 10~\mum\ peak is often too  strong.
The laboratory silicates produce the opposite problem:  the 
18~\mum\ feature is often too strong compared to the 10~\mum\ 
feature and peaks at too short of a wavelength.  The 10~\mum\ 
feature peaks at slightly too short of a wavelength as well.
In the 19 cases where the laboratory silicate provided the 
best fit, the model with the maximum considered value of 5\% 
metallic iron worked best in 17 cases.  Kemper et al.\ (2002) 
used 4\% in their study of a Galactic OH/IR star.  The use of 
metallic iron is linked to the fact that laboratory silicates 
like olivine provide too little opacity in the near-IR, a 
notion that goes back to the introduction of the term ``dirty 
silicates'' by Jones \& Merrill (1976).

The opacity of the laboratory silicates has been calculated for small
particles and for olivine and aluminium oxide for single-sized
spherical grains.  (For iron we followed the procedure of Kemper et
al.\ to use a CDE.)  The opacity of the mixture was calculated by
adding the absorption coefficients of the components in proportion.
The true composition and shape of the dust particles is immensely 
more complex than adopted here.  One could consider the effects of
ellipsoidal grains, CDE distributions, core-mantle grains, 
multi-layered grains, or even fluffy aggregates.  It is likely
that shape and size distribution and composition will all vary with
distance from the central star, and may even vary with the phase of
the pulsation cycle.  We will briefly address some of these effects
below, by comparing different models to observations of HV 12793.
This source has an SED which is very well fitted, 
an optically thin shell, and is
best fit with a mixture of laboratory silicates with 5\% iron and
95\% olivine.  This mixture is typical for our best-fitting
laboratory silicate models.

\begin{figure}[!t] 

\begin{minipage}{0.49\textwidth}
\resizebox{\hsize}{!}{\includegraphics[angle=-90]{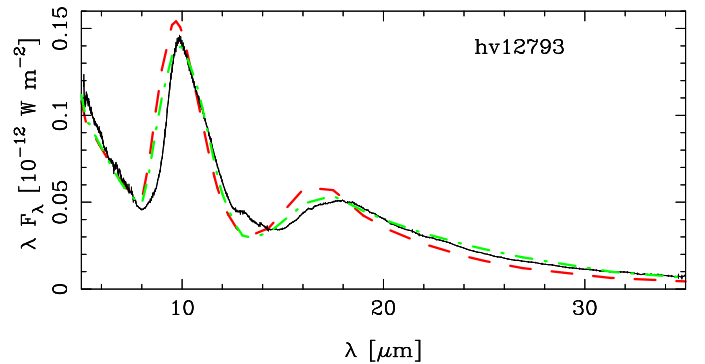}}
\end{minipage}

\caption[]{ 
The effect of grain shape, showing the original best-fitting 
model with small spherical olivine grains (red dashed line) and 
with a CDS (green dot-dashed line), both with 5\% iron.  Models 
assuming a CDE or DHS give results nearly identical to the CDS model.
}
\label{Fig-org-cds} 
\end{figure}

Figure~\ref{Fig-org-cds} shows the effect of different shapes, 
calculated for small particles, that have analytical solutions for 
the absorption properties (see Min et al.\ 2003): CDE, continuous 
distribution of spheroids (CDS), and a distribution of hollow 
spheres (DHS).  In the last case, the optical properties are 
averaged over a uniform distribution in volume fraction between 
0 and $f_{\rm max}$ of a vacuum core, where the material volume 
is kept constant.  While CDE and CDS models probe the effect of 
deviations from homogeneous spherical grains, DHS models probe 
the effect of porosity.  As Min et al.\ showed, the simple 
core-mantle approach in DHS is essentially equivalent to the 
general case of random vacuum inclusions according to effective 
medium theory (EMT).  
In the models, $T_{\rm c}$ was fixed and luminosity and MLR
were fitted for grains with 3, 4 and 5\% iron.
It turns out that CDE, CDS and DHS models show very similar 
results (with CDS models producing the best fits, formally).
In all cases the best-fitting models have 5\% iron.
Figure~\ref{Fig-org-cds} shows the best fit with small spherical 
grains and with a CDS (both with 5\% iron).  Using a CDS (or CDE
or DHS) improves the fit to the features at 10 and 18~\mum.

\begin{figure}[!t] 

\begin{minipage}{0.49\textwidth}
\resizebox{\hsize}{!}{\includegraphics[angle=-90]{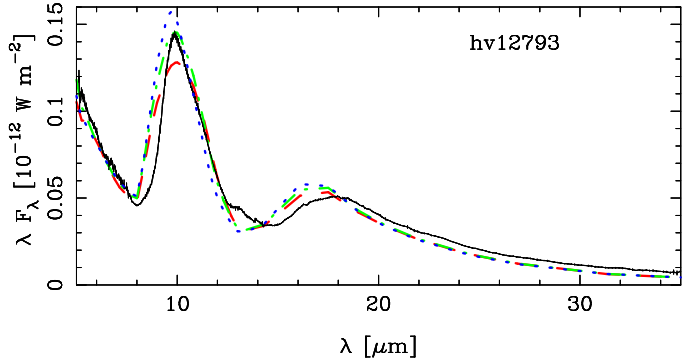}}
\end{minipage}

\caption[]{ 
The effect of grain size for homogeneous spheres (95\% olivine 
and 5\% iron).  Grain sizes are 1.5 (red dashed line), 1.2 
(green dot-dashed line) and 0.6~\mum\ (blue dotted line). 
Grains smaller than 0.6~\mum\ are indistinguishable from the 
0.6-\mum\ case.
}
\label{Fig-dhs0} 
\end{figure}

Figure~\ref{Fig-dhs0} shows the effect of different grain size, 
for a DHS with $f_{\rm max}$= 0 (homogeneous spheres).  It shows 
that grain size also has an effect on the fit to the 10--18~\mum\ 
region and that grains with a size around 1~\mum\ fit reasonably well.

\begin{figure} 

\begin{minipage}{0.49\textwidth}
\resizebox{\hsize}{!}{\includegraphics[angle=-90]{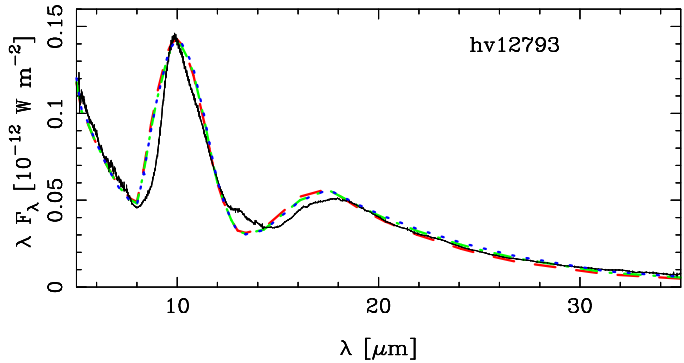}}
\end{minipage}

\caption[]{ 
The effect of porosity using a DHS with $f_{\rm max}$= 0.4 
(red dashed line), 0.7 (green dot-dashed line) and 0.9 
(blue dotted line) for a grain size of 1.2~\mum.
}
\label{Fig-dhs1} 
\end{figure}

Figure~\ref{Fig-dhs1} shows the effect of porosity, with
$f_{\rm max}$= 0.4, 0.7 and 0.9 for a grain size of 1.2~\mum.  
Min et al.\ (2005, 2007) show that a DHS with $f_{\rm max}$= 0.7
provides a good fit to the shape of the interstellar silicate 
feature at 10~\mum, while with a value of 0.4 they can reproduce 
the polarisation properties of quartz measured in laboratory data. 
The effect of increasing the porosity on the spectra does not appear
to be very large, mostly increasing the opacity at long wavelengths, 
so the value of $f_{\rm max}$ is not well constrained from the present data.
Luminosity and MLR were minimised in all of these calculations, and
the best-fitting values do change: from $L$= 118~000 \lsol\ and \mdot=
5.1 $10^{-7}$ \msolyr\ for $f_{\rm max}$= 0.4 to $L$= 136~000
\lsol\ and \mdot= 3.3 $10^{-7}$ \msolyr\ for $f_{\rm max}$= 0.9.

We conclude that, compared to the small-particle limit, a
distribution of shapes shifts the peak of the 9.8 and 18~\mum\
feature to slightly longer wavelengths, in better agreement with
observations.  Assuming porous grains achieves the same effect.
When considering grains of finite size and a DHS we find for HV 12793
that a grain size of around 1~\mum\ fits the observations.  This
result is typical of many of the M giants in the sample, although the
porosity is not well constrained.

H\"ofner (2008) recently studied the winds of M giants using 
a dynamical atmosphere and wind models and concluded that 
radiation pressure is sufficient to drive the outflow if the 
restriction of small particles is relaxed.  She also showed 
that grain growth naturally halts at particle sizes 
of about 1~\mum.  The present study provides observational
support to these theoretical conclusions.

\subsection{Dust in the C-stars}

\subsubsection{Silicon carbide}
\label{sec-c}

While amorphous carbon dominates the dust produced by carbon
stars (e.g.\ Martin \& Rogers 1987), silicon carbide (SiC) 
produces the most recognisable dust emission feature at 
$\sim$11.3~\mum\ (Treffers \& Cohen 1974; and references
therein).  The models here utilise the SiC optical constants
of Borghesi et al.\ (1985) and P\'{e}gouri\'{e} (1988).  In 
reality, more complex situations may occur than assumed in 
the standard model, namely the coexistence of two grains (SiC 
and AMC) at the same temperature (adding the absorption 
coefficients in proportion) in the small particle limit.

\begin{figure} 

\begin{minipage}{0.49\textwidth}
\resizebox{\hsize}{!}{\includegraphics[angle=-90]{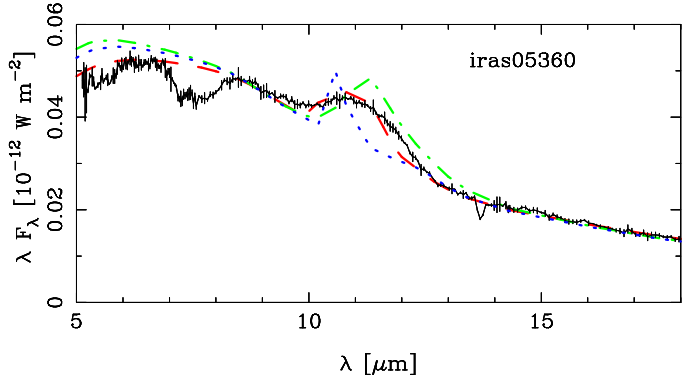}}
\end{minipage}

\caption[]{ 
A comparison of different types of SiC assuming small spherical particles:  
8\% $\beta$-SiC from Borghesi et al.\ (1985) (red dashed line), 
8\% $\alpha$-SiC from P\'{e}gouri\'{e} (1988) (green dot-dashed line), and 
2\% $\beta$-SiC from Pitman et al. (2008) (blue dotted line), 
with the remaining fraction of the dust being AMC.
The models are tied to the observed spectrum based on the 
flux density in the 8.5--9.5~\mum\ region.
}
\label{Fig-3sics} 
\end{figure}

Some of the possible effects are illustrated here, using 
IRAS 05360$-$6648 as the test case.
Figure~\ref{Fig-3sics} compares the best fit using 8\% 
$\beta$-SiC from Borghesi et al. (1985), 8\% $\alpha$-SiC 
from P\'{e}gouri\'{e} (1988), and 2\% $\beta$-SiC from 
Pitman et al.\ (2008), assuming small spherical particles.  
The constants from Borghesi et al.\ fit the data best; the
other two are too sharply peaked compared to the 
observations.

\begin{figure} 

\begin{minipage}{0.49\textwidth}
\resizebox{\hsize}{!}{\includegraphics[angle=-90]{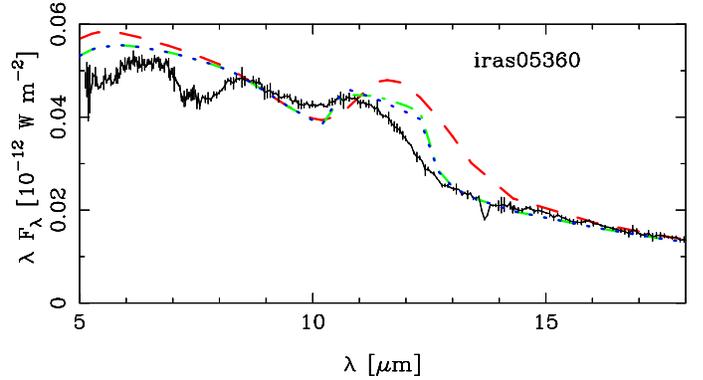}}
\end{minipage}

\caption[]{ 
The effect of shape distribution: 
CDE with 6\% $\beta$-SiC from  P\'{e}gouri\'{e} (1988) (red dashed line), 
CDE (green  dot-dashed line), and  
CDS (blue dotted line) with 2\% $\beta$-SiC from Pitman et al.\ (2008), 
with the remaining fraction being AMC. 
}
\label{Fig-71518} 
\end{figure}

Figure~\ref{Fig-71518} shows the effects of the CDE and CDS 
approximations.  Since the SiC from P\'{e}gouri\'{e} (1988) 
peaks to the right of that of Pitman et al.\ (2008) this 
effect is enhanced when using a distribution of shapes, and 
the latter clearly provides the better fit.  On the other 
hand, the shape of the feature becomes ``boxy'' which is not observed.

\begin{figure} 

\begin{minipage}{0.49\textwidth}
\resizebox{\hsize}{!}{\includegraphics[angle=-90]{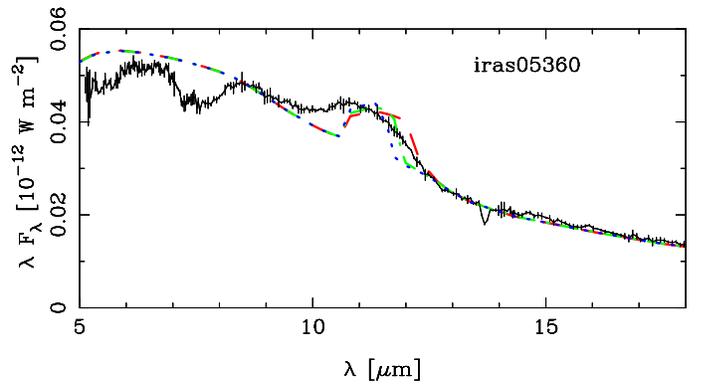}}
\end{minipage}

\caption[]{ 
The effect of using a DHS with $f_{\rm max}$ equal to
0.9 (red dashed line), 0.8 (green dot-dashed line), and  
0.7 (blue dotted line), with  1\% $\beta$-SiC from 
Pitman et al.\ (2008) and 99\% AMC.
}
\label{Fig-dhssic} 
\end{figure}

Figure~\ref{Fig-dhssic} shows the effect of using DHS with 
different values of $f_{\rm max}$ in the case of $\beta$-SiC 
from Pitman et al.\ (2008).  The model with $f_{\rm max}$ = 
0.8 fits the data reasonably well and slightly better than 
the CDE and CDS approximation, but still there is a lack of 
emission on the blue side of the feature.

\begin{figure} 

\begin{minipage}{0.49\textwidth}
\resizebox{\hsize}{!}{\includegraphics[angle=-90]{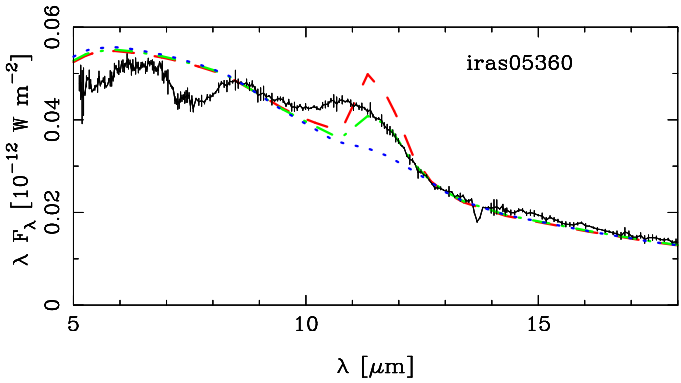}}
\end{minipage}

\caption[]{ 
The effect of using spherical core-mantle grains, with a SiC 
core of 0.01 $\mu$m, and a total radius of 
0.03 (red dashed  line), 
0.04 (green dot-dashed line), and 
0.08 (blue dotted  line).  
The volume fraction of SiC in these grains is $\sim$4, 2, and 0.2\%, respectively.
}
\label{Fig-cm} 
\end{figure}

Kozasa et al.\ (1996), and more recently Papoular (2008), 
considered grains with a core of SiC and a mantle of AMC, as 
an explanation for the observed variations in the strength
of the SiC feature.  Figure~\ref{Fig-cm} shows the 
best-fitting models for grains with a 0.01~\mum\ SiC core, 
and a total radius of 0.03, 0.04, 0.08~\mum.  The figure 
shows that changing the volume fraction of the SiC core from 
about 4 to 2 to 0.2\% changes strength of the SiC feature 
from stronger than in any spectrum in the present sample to 
effectively zero.  In the test calculation by Papoular, the 
grain reached its final size about one year after its initial
condensation, with a core volume of 12\%, which would produce
a SiC feature stronger than any in the sample considered here.
Assuming an expansion velocity of 10~\ks, a dust particle 
would travel about 1 stellar radius per year in the case of 
IRAS 05360$-$6648.  The formation of the coating must 
therefore proceed much more efficiently than in the test 
calculation of Papoular for core-mantle grains to be a viable 
option for explaining the maximum observed strength of the 
SiC feature.  Speck et al.\ (2009) note only one case where 
presolar SiC grains are found in the cores of carbon grains 
in meteoritic samples, adding further doubt about the 
importance of grains with SiC cores and AMC mantles.

Lagadec et al.\ (2007) argue for a SiC $\rightarrow$ C condensation
sequence in Galactic C-stars and Leisenring et al.\ (2008, their
figure~13) discuss three condensation sequences, two of which start
with SiC cores.  The absence of a very strong SiC feature in any
Galactic and MC C-star predicted for a core-mantle grains with even
moderate SiC volume fraction appears to rule out condensation
sequences {\sc ii} and {\sc iii} in Leisenring et al.\ (2008).
Leisenring et al. also argue that SiC and C condense
near-simultaneously.  In that case a mix of condensation sequences
{\sc i} (SiC on top of C) and {\sc ii} could take place.

\subsubsection{Magnesium sulfide}
\label{secMGS}

In their sample Leisenring et al.\ (2008) found 6 out of 29 (21\%)
C-stars in the SMC to show MgS, and 27/44 = 61\% stars in the LMC (and
14/34 = 41\% for Galactic sources). In our larger sample the ratio for
the SMC now becomes 8/33 = 24\%, and 48/68 = 71\% for the LMC.

Only one of the 22 stars with $R_{\rm outer} < 50 R_{\rm inner}$ show MgS 
(The exception is MSX LMC 1213 with $R_{\rm outer} = 41 R_{\rm inner}$.)  
The dust temperature at the outer radius ranges from 175 to 400 K in
these stars, suggesting that MgS forms at temperatures below 400 K,
consistent with the MgS temperatures derived by Hony et al.\ (2002).

Recently, Zhukovska \& Gail (2008) studied the condensation 
of MgS in the outflows of C-stars using thermo-chemical models 
including grain growth.  
They preferred a scenario where MgS mantles grew on SiC cores.
Unfortunately, they did not consider the case of SiC+AMC+MgS.
However,
as MgS forms at the lowest temperatures, one could expect to have
grains consisting of a SiC core, an inner mantle of AMC (in spite of
the remarks in the previous section) and an outer mantle of MgS.
Alternatively, a model with pure AMC grains on the one hand and
grains with a MgS mantle on a core of SiC on the other hand are considered below.
Some test calculation are done for such grains\footnote{Adapting the code for n-layered spherical grains available at 

http://www.astro.spbu.ru/JPDOC/CODES/NMIE/n-miev3a.for}.
Our model does not permit a gradient in grain properties 
within the circumstellar dust shell; thus we cannot account 
for the possibility that grains might have an outer layer of
MgS only in the outerparts of the shell.
We remind the reader that the presence of MgS may mask the SiC
strength, making our estimates of the SiC mass fraction a lower
limit in spectra showing MgS emission.

\begin{figure} 

\begin{minipage}{0.49\textwidth}
\resizebox{\hsize}{!}{\includegraphics[angle=-90]{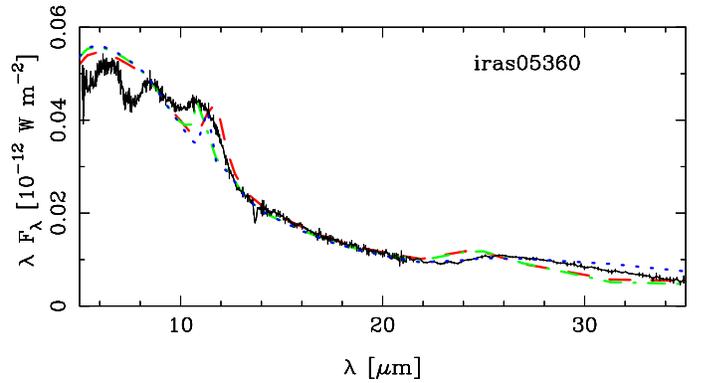}}
\end{minipage}

\caption[]{ 
Some models that include MgS.
(1) Core with 2\% SiC, inner mantle with 91\% AMC, outer 
mantle with 7\% MgS.  CDE using the optical constants 
calculated using EMT (red dashed line), 
(2) two grain model, 90\% AMC and 10\% of a core-mantle 
grain with 45\% SiC core and 55\% MgS mantle (green 
dot-dashed line), and  
(3) two grain model, 94\% AMC and 6\% of a core-mantle grain 
with 3\% SiC core and 97\% MgS mantle, assuming CDE using the 
optical constants calculated using EMT (blue dotted line).
}
\label{Fig-mgs} 
\end{figure}

Hony et al.\ (2002) showed the importance of grain shape in 
fitting MgS to the observed 30~\mum\ feature.  They found 
that a CDE provided a good fit.  Here we also consider CDE, 
and in this respect one would not favour core-mantle grains 
with MgS+SiC+AMC (see Fig.~\ref{Fig-mgs}). 
The Mie-code employed also provides the optical constant of 
the aggregate following EMT, and we used this to calculate 
the absorption coefficients in a CDE.  As MgS is a minor 
species, the effect of broadening the MgS feature as seen in 
the models of Hony et al.\ does not occur.  In this respect, 
AMC as one grain, with a separate core-mantle grain of 
SiC+MgS fits the data better.  The volume fraction of MgS in 
such a grain is substantial, and using the optical constants 
of the aggregate in the CDE approximation does broaden
the feature sufficiently.

\subsection{Mass-loss and stellar evolution}
\label{sec-ML}

This section compares the derived MLRs and other quantities by
comparing the observations with the AGB models of Vassiliadis \& Wood
(1993, hereafter VW) and synthetic models based on recipes developed
by Wagenhuber \& Groenewegen (1998, hereafter WG), tuned to reproduce
the VW models, as explained in Appendix C of Groenewegen \& Blommaert (2005).  
This last option allows us to investigate initial masses and MLRs
different from those available from the VW models alone (with a
maximum initial mass of 5~\msol).
From VW we take the models with $Z$=0.008 as representative 
of the LMC, since they show qualitatively similar behaviour. 
Using the WG implementation of the VW models, we have 
calculated a model for 7.9 \msol, which is the largest mass 
for which the WG models converge.
In addition we have used the WG recipes to calculate models 
using the Reimers law with a scaling factor of 5 (hereafter 
the Reimers models), which Groenewegen \& de Jong (1993) 
advocated to fit many of the observables of C stars in the LMC.
The major divergence is the recipe to determine the 
MLR on the AGB.  In the case of the Reimers law this is 
\mdot\ $\sim L \, R /M$, while in the case of VW it is 
basically the minimum of the single scattering limit 
\mdot $= 2.02 \, 10^{-8} \, L / v_{\rm exp}$, and an 
empirical relation between  $\log \dot{M}$ and $P$.
The (fundamental mode) period, $P$, is calculated from a 
period-mass-radius relation (see VW for details).

\begin{figure*}
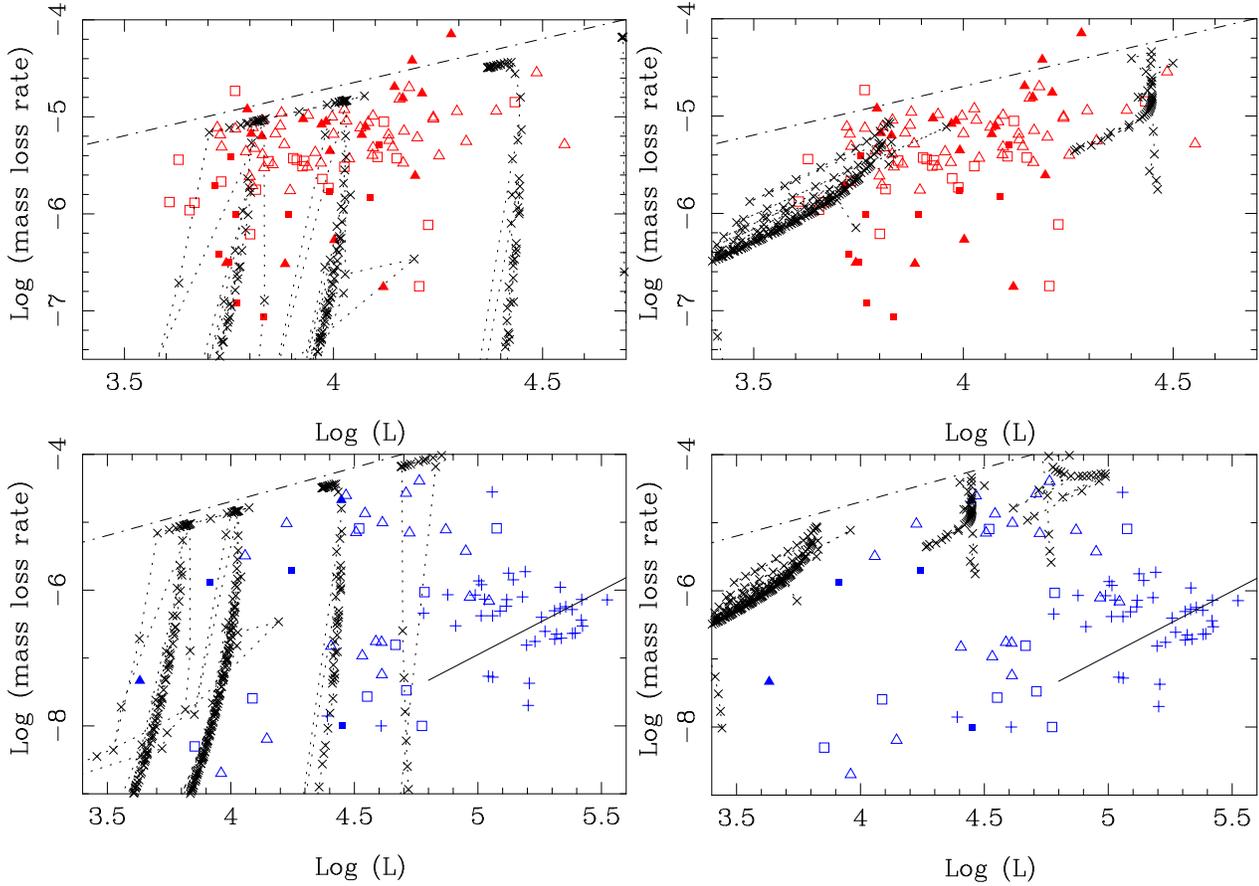
 

\begin{minipage}{0.45\textwidth}
\resizebox{\hsize}{!}{\includegraphics{ML_L_VW_PAPER.ps}}
\end{minipage}
\begin{minipage}{0.45\textwidth}
\resizebox{\hsize}{!}{\includegraphics{ML_L_WG_PAPER.ps}}
\end{minipage}

\caption[]{ 
Mass-loss rate versus luminosity for C stars (top panel, red colours),
M stars (bottom panel, blue colours), VW models (left), and Reimers models 
(right).
AGB stars in the SMC are plotted as squares, in the LMC as 
triangles.
Objects with Mira like pulsation amplitudes are plotted with 
open symbols, objects with smaller amplitudes as filled 
symbols.
RSG are plotted as plus-signs independent of host galaxy and 
pulsation amplitude.
The VW and Reimers models are plotted as crosses connected by 
the dotted line for initial masses of 1.5, 2.5, 5.0 and 
7.9~\msol, but not every track is visible in every panel. 
Each cross represents a time interval of 5000 years.
The dot-dashed line indicates the single scattering limit for 
a velocity of 10~\ks.
The solid line is the relation found by Verhoelst et al.\ (2009) for Galactic RSG.
} 
\label{Fig-MLLum} 
\end{figure*}

Figure~\ref{Fig-MLLum} shows the relation between MLR and
luminosity, with the VW and Reimers model tracks overplotted 
(the crosses connected with the dotted lines).  From the 
individual evolutionary tracks, a model is plotted every 5000 
years.  The density of points is therefore representative for 
the time spent at the certain position in the diagram.  It 
also explains the ``excursions'' which are due to the finite 
probability of catching a star during a thermal pulse or 
during the ``luminosity dip'' that follows.  Models are 
plotted for initial masses of 1.5, 2.5, 5.0 and 7.9~\msol, 
which evolve at increasing luminosity.  The Reimers 
1.5 \msol\ model is only identified by the few points near 
$\log L$= 3.5 \lsol\ and MLRs below 10$^{-7}$ \msolyr.

Several conclusions can be drawn.  From the observational 
point of view, the MLRs of the C stars in the sample are 
located in a strip with a width of about 1 dex, slightly
lower than the single-scattering limit.  At the lowest MLRs,
the majority of stars have smaller amplitudes (filled 
symbols).  The dot-dashed line indicates that the 
single-scattering limit for a velocity of 10 \ks\ which is 
the same velocity as adopted in the RT models.  However 
there is an additional systematic uncertainty in the 
$y$-scale because of the adopted dust-to-gas ratio in the RT  modelling.

The MLR distribution for the O-rich AGB stars is less clear.  
One could describe two sequences, one of low MLRs
(10$^{-7}$-10$^{-6}$ \msolyr) for $\log L \approx 3.7-4.7$ 
and one of large MLRs (3 10$^{-6}$- 3 10$^{-5}$ \msolyr for 
$\log L \approx 4.1-4.8$), which is roughly consistent with 
the VW models, as discussed below. 

The MLRs of the RSG scatter around the relation of Galactic
RSG recently derived by Verhoelst et al. (2009).

The comparison to the evolutionary models is interesting, and 
favours overall the mass-loss recipe adopted in VW.  Taking M 
and C stars together it is clear from Fig.~\ref{Fig-MLLum} 
that there is a large scatter in MLR for a given luminosity.  
This is clearly not predicted by the Reimers models, where 
the luminosity fixes the MLR.
The VW models do predict a large variation of MLR at a
given luminosity, as observed, and consistent with other 
evolutionary considerations.  The majority of C stars are 
bound between the 1.5- and 5.0-\msol\ models, which indeed 
is the range where one believes that C-stars form in the 
LMC (Groenewegen \& de Jong 1993).  
The lowest luminosity C-stars in the sample are located in 
the SMC (the open squares) where the initial mass to become 
a C star is slightly lower then in the LMC (Groenewegen 1993). 
The VW models also explain that, in the 1.5--5-\msol\ 
mass range, the M stars have lower MLRs than the C stars at 
earlier times in their evolution.

There are essentially no C stars brighter than $\log L = 4.5$, but
many M stars are brighter than that, and they span a large range in
MLRs. The separation near 5~\msol\ is thought to be due to HBB (Smith
et al. 1995).  The 7.9\msol\ model based on the VW mass-loss recipe
nicely passes through the data.

Although the VW models generally provide a qualitatively 
correct picture of mass-loss evolution along the AGB for both 
M  and C stars, they fail on one point.  If the sample under 
study were complete, then the distribution of observed data 
should match the distribution of the crosses, but this is
clearly not the case.  
While the observations give a fairly uniform distribution of MLRs at a
given luminosity for the C-stars, the VW models predict that most of
the MLRs are in fact limited by the single-scattering limit.  
As in VW the expansion velocity is assumed to be a function of period.
With a maximum of 15~\ks, the maximum MLR in the VW is slightly lower
than the dot-dashed line which represents the single-scattering limit
for 10~ \ks.

One could argue that this a a selection effect and that the present
sample is incomplete for the low MLRs and lowest luminosities, which
would predominantly be O-rich stars.
It was verified however, by comparing the cumulative absolute
magnitude distribution function of the C-stars in this sample to the
sample of about 1800 C-stars selected on colour from the entire SAGE
survey by Matsuura et al. (2009), and using the relation in
Fig.~\ref{Fig-BC} to transform colours to bolometric magnitudes, that
the present C-star sample is unbiased in absolute magnitude down to 
$M_{\rm bol} \approx -5.5$ ($\log L$= 4.0).
\bigskip

\begin{figure*}  

\begin{minipage}{0.45\textwidth}
\resizebox{\hsize}{!}{\includegraphics{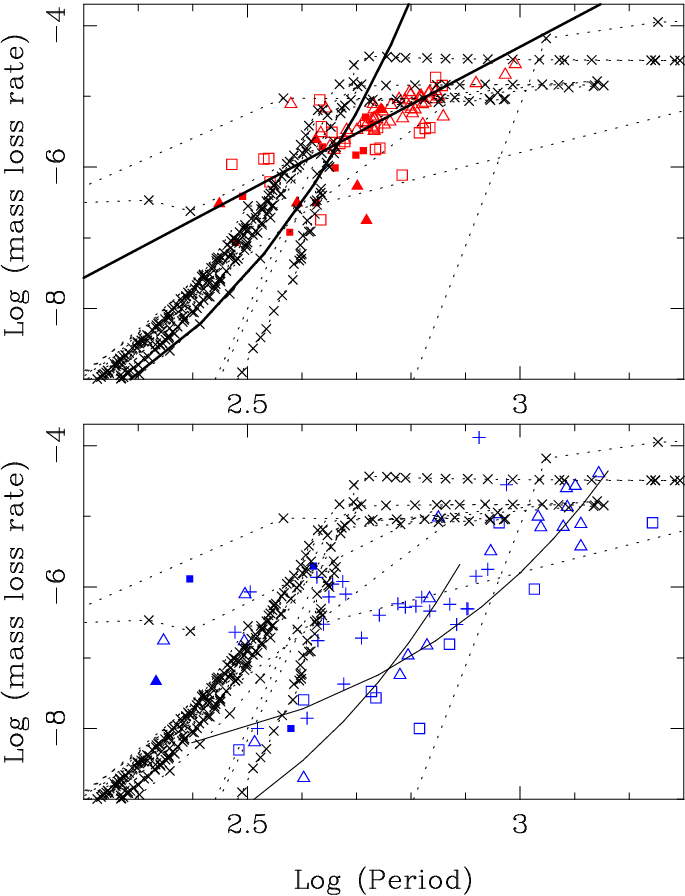}}
\end{minipage}
\begin{minipage}{0.45\textwidth}
\resizebox{\hsize}{!}{\includegraphics{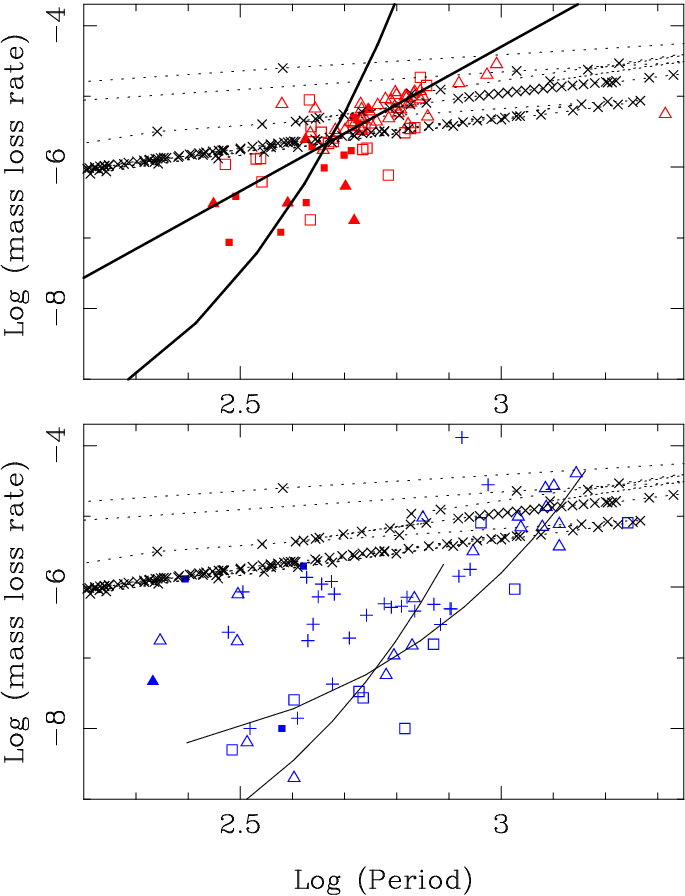}}
\end{minipage}

\caption[]{ 
Mass-loss rate versus period for C stars (top panel, red colours), 
M stars (bottom panel, blue colours), VW models (left), and Reimers 
models (right).
Symbols are as in Fig.~\ref{Fig-MLLum}.
In the upper panels, the curved solid line indicates the 
MLR-period relation from VW, the straight solid line the 
relation derived for Galactic C-rich Miras by Groenewegen 
et al.\ (1998).
In the bottom panel, the two lines indicate a fit to most of 
the M stars, and the VW relation for a period 0.6 times the 
pulsation period.
}
\label{Fig-MLPer} 
\end{figure*}

Figure~\ref{Fig-MLPer} shows the derived MLR plotted against 
pulsation period, for the C stars and M stars, with the VW
and Reimers models overplotted.  Overall, the VW models cover 
the area occupied by the C stars, except in the early phases 
of evolution when the stars are not yet carbon-rich.
The Reimers models do not cover the range in mass-loss of the 
C stars.  It is surprising that the M stars do not follow the 
VW relation, even though it was derived for a sample that 
included O-rich Mira variables.  A relation similar to Eq.~2 
of VW that fits most of the O-rich data is 
$\log \dot{M}= -9.0 + 0.0032 P$.
An alternative interpretation is that the largest MLRs of the 
M stars are consistent with the VW model, namely that they 
are on the horizontal part of the most massive evolutionary 
tracks, and that the lower MLRs are roughly consistent with a 
track \it parallel \rm to the VW relation for a period roughly 
0.6 times the pulsation period.  A connection with overtone 
pulsation comes to mind although the amplitudes of these 
M stars indicate Mira-like (fundamental-mode) pulsations.
\bigskip

\begin{figure}  

\begin{minipage}{0.49\textwidth}
\resizebox{\hsize}{!}{\includegraphics{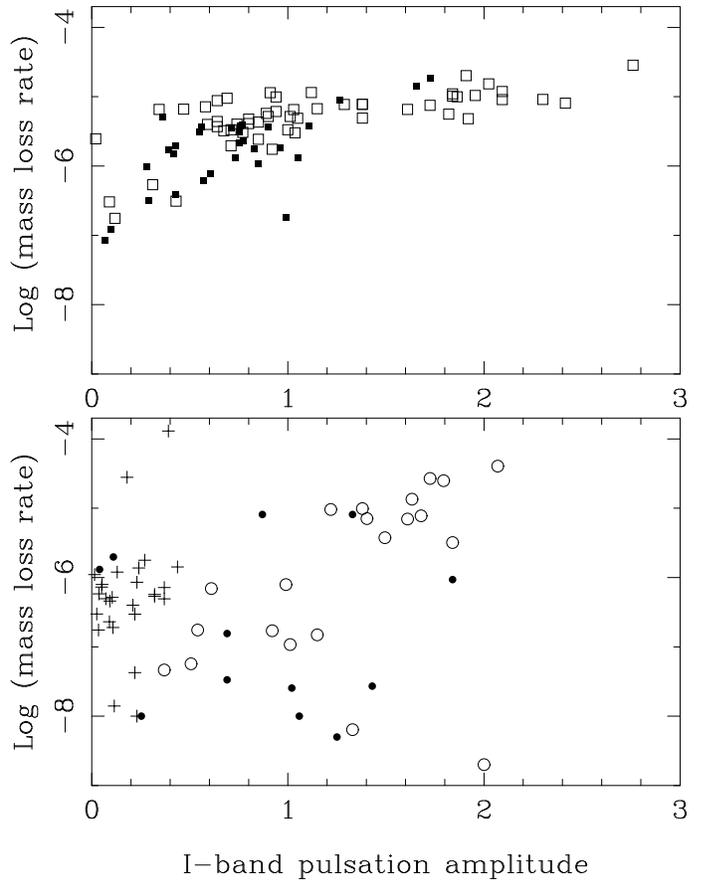}}
\end{minipage}

\caption[]{ 
Mass-loss rate versus (estimated or measured) $I$-band 
pulsation amplitude for C stars (top panel) and M stars 
(bottom panel).
C stars in the LMC are plotted as open squares, in the SMC 
as filled squares.
%
O-rich AGB stars in the LMC are plotted as open circles, and in 
the SMC as filled circles.  RSG are plotted as plus signs.
} 
\label{Fig-MLAmp} 
\end{figure}

Figure~\ref{Fig-MLAmp} plots the MLR versus the observed or
estimated pulsation amplitude in the $I$-band.  Whitelock et 
al.\ (2003) showed a similar diagram with MLR vs.\ the 
$K$-band.  For the larger amplitudes there is a reasonably 
well defined and almost flat relation. Outside the regime 
of Mira-like pulsation there is much more scatter.
For the M stars there may be a trend of larger MLRs with 
larger amplitudes that is steeper than for the C stars but 
there is even more scatter.
\bigskip

\begin{figure}  

\begin{minipage}{0.49\textwidth}
\resizebox{\hsize}{!}{\includegraphics{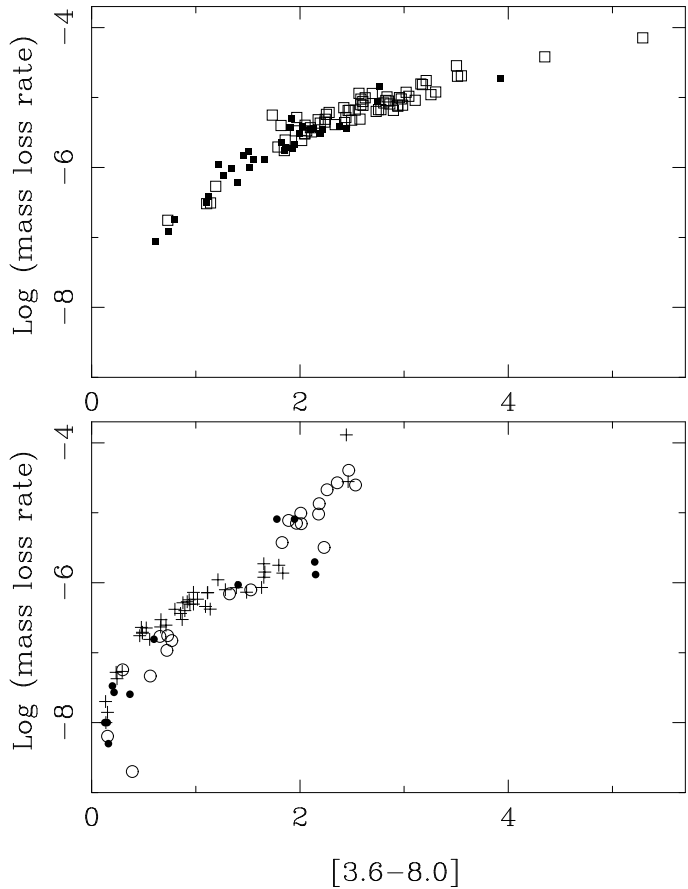}}
\end{minipage}

\caption[]{ 
Mass-loss rate versus colour for C-stars (top panel), and M-stars (bottom panel).
Symbols as in Fig.~\ref{Fig-MLAmp}.
} 
\label{Fig-MLCol} 
\end{figure}

Figure~\ref{Fig-MLCol} plots the MLR as a function of 
[3.6]$-$[8.0] colour.  Whitelock et al.\ (2003) presented
a similar diagram with MLR vs.\ [K]$-$[12].  Generally, 
redder colours are associated with larger MLR, as expected. 
For C stars the relation is tight, with no dependence on
metallicity apparent (assuming that the expansion velocity 
and dust-to-gas ratio are on average the same in the LMC 
and SMC).

For the M stars there appear to be two relations, with both
RSGs and AGB stars following the same relation up to a 
certain colour, with a discontinuity for stars that are 
associated with the final stages of evolution of the most 
massive intermediate stars according to the VW models (see 
Fig.\ref{Fig-MLLum}).
Together with the relations in Fig.~\ref{Fig-BC}, the 
[3.6]$-$[8.0] colour can be used to estimate luminosity and MLR.
%




%
%
 
\bigskip

\begin{figure}  

\begin{minipage}{0.49\textwidth}
\resizebox{\hsize}{!}{\includegraphics{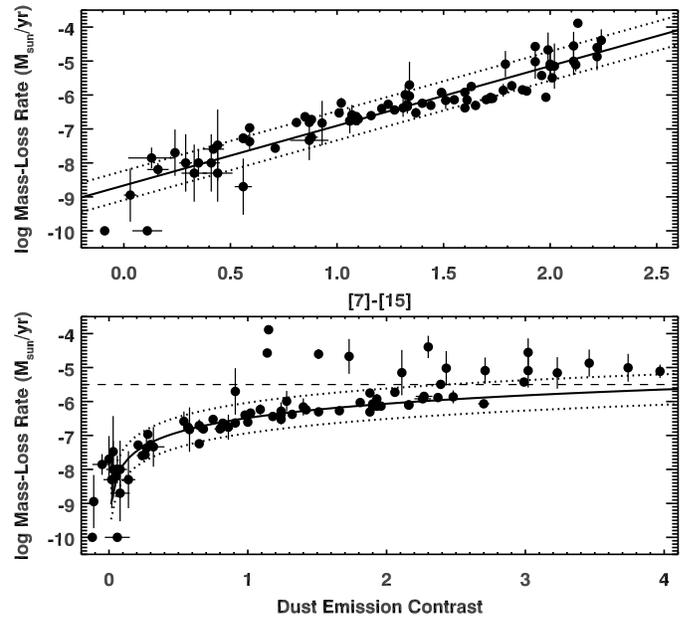}}
\end{minipage}

\caption[]{ 
%
%
%
Mass-loss rate as a function of [7]$-$[15] color (top panel)
and DEC (dust emission contrast; bottom panel).  In both
panels, the solid line plots the functions given in equations~\ref{eq-xx}
 and ~\ref{eq-yy}, and the dotted lines show the average uncertainty,
which is formally the standard deviation of the difference
between the actual data and the fitted function.  The dashed
line in the bottom panel separates the data excluded from the
fit.  Also excluded were the four points with negative DECs.
Neither panel includes foreground objects.
} 
\label{Fig-MLDEC} 
\end{figure}

Sloan et al.\ (2008) presented two tools to measure the amount
of dust in their sample of evolved oxygen-rich Magellanic
stars.  The first, dust emission contrast (DEC) was originally
introduced by Sloan \& Price (1995), and is the ratio of the
dust emission to stellar emission from 7.67 to 14.03~\mum,
based on a stellar photosphere fitted to the spectrum over
the range 6.8--7.4~\mum\footnote{Sloan et al.\ (2008) shifted
this range for the IRS data from the range originally defined
for the LRS data.}.  The second was the [7]$-$[15] color,
integrated at 6.8--7.4 and 14.4--15.0~\mum, wavelength ranges
chose to avoid molecular absorption from H$_2$O at 6.6~\mum\
and SiO, starting at 7.5~\mum, and dust emission features from
silicates and alumina.  Figure~\ref{Fig-MLDEC} plots the
mass-loss rates derived here with these quantities obtained
from the IRS data.  The mass-loss rates follow the [7]$-$[15]
color closely:
\begin{equation}
   \log \dot{M} = 1.759 \; ([7]-[15]) - 8.664,
\label{eq-xx}
\end{equation}
with an average uncertainty of 0.43 dex, or 27\% of the
mass-loss rate.  Thus, the [7]$-$[15] color provides an
excellent means of estimating mass-loss rates from infrared
spectra, provided the dust is configured in an outflowing
shell as opposed to a disc.

Figure~\ref{Fig-MLDEC} also plots the mass-loss rate as a
function of DEC, and the relation is more complex than with
[7]$-$[15] color.  The DEC was originally defined for
optically thin dust shells, and it breaks down as a useful
measure when the dust grows optically thick.  Once mass-loss
rates exceed $\sim$10$^{-5.5}$ \msolyr, the 10~\mum\ silicate
emission feature will begin to self-absorb, and the DEC will
begin to decrease, even as the mass-loss rate grows.  The
[7]$-$[15] color is thus a more robust measure.  There is
still a relation between mass-loss rate and DEC, provided
certain caveats are kept in mind.  Fitting the data where
the DEC is positive and the mass-loss rate is less than
10$^{-5.5}$ \msolyr\ gives the relation
\begin{equation}
  \dot{M} =  3.281~10^{-7} ({\rm DEC} )^{1.392}
\label{eq-yy}
\end{equation}
with an average uncertainty of 0.45 dex, or 28\% of the
mass-loss rate.  As Figure~\ref{Fig-MLDEC} shows, though,
the data do not fit this relation well when the DEC
exceeds $\sim$2.5.

One must keep in mind that these mass-loss rates assume
a gas-to-dust ratio of 200, since that is the value assumed
in the radiative transfer modelling.  The spectroscopy really
measures the dust, so it would be appropriate to scale these
relations accordingly for different gas-to-dust ratios.

\bigskip

\begin{figure*}
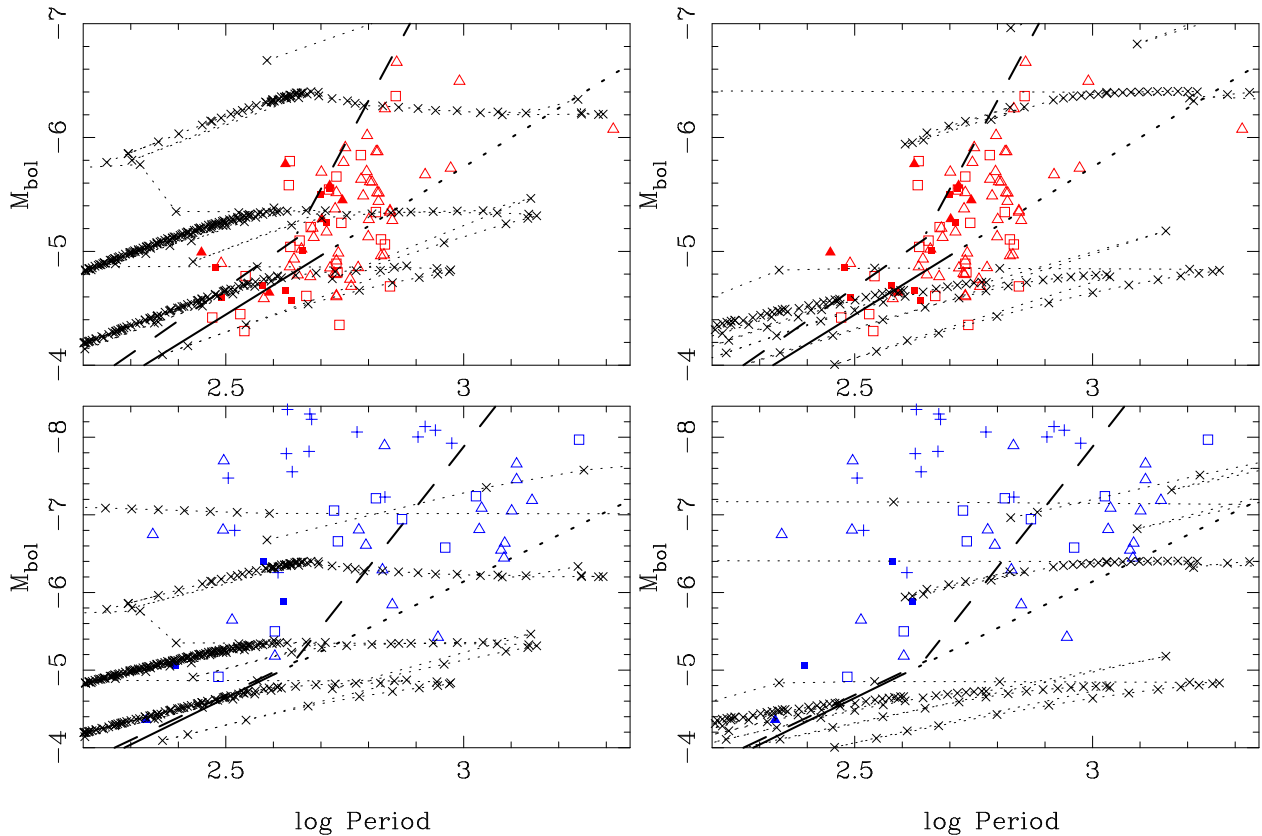
 

\begin{minipage}{0.45\textwidth}
\resizebox{\hsize}{!}{\includegraphics{MbolPer_VW_PAPER.ps}}
\end{minipage}
\begin{minipage}{0.45\textwidth}
\resizebox{\hsize}{!}{\includegraphics{MbolPer_WG_PAPER.ps}}
\end{minipage}

\caption[]{ 
Bolometric magnitude versus pulsation period for C-stars (top panel, red colours), 
and M-stars (bottom panel, blue colours), and for VW models (left) and Reimers models (right).
AGB stars in the SMC are plotted as squares, in the LMC as triangles.
Objects with Mira like pulsation amplitudes are plotted with open symbols, and filled symbols otherwise.
RSG are plotted as plus-signs independent of host galaxy and pulsation amplitude.
For the C-star panels, the solid line indicates the $PL$-relation from
Groenewegen \& Whitelock (1996) and the dotted line the extrapolation beyond $\log P$= 2.7.
For the O-star panels,  the solid line indicates the $PL$-relation from
Feast et al. (1989) and the dotted line the extrapolation beyond $P$= 400 days.
The dashed line with the break and discontinuity at $\log P$= 2.65 in both panels is
the $PL$-relation from Hughes \& Wood (1990) based on a sample dominated by O-stars.
}
\label{Fig-MbolPer} 
\end{figure*}

Figure~\ref{Fig-MbolPer} shows the bolometric $PL$-relation with the
VW and Reimers models overplotted, together with various
(extrapolated) $PL$-relations for C- and O-stars (Feast et al. 1989,
Hughes \& Wood 1990, Groenewegen \& Whitelock 1996).  

Most of the VW points are to the left of the data points, which can be
interpreted as meaning that pulsation in the fundamental period is
only reached towards the end of AGB evolution, which is consistent
with the findings of Lebzelter \& Wood (2005, 2007) for the clusters
47 Tuc and NGC 1846.

\begin{figure} 

\begin{minipage}{0.49\textwidth}
\resizebox{\hsize}{!}{\includegraphics{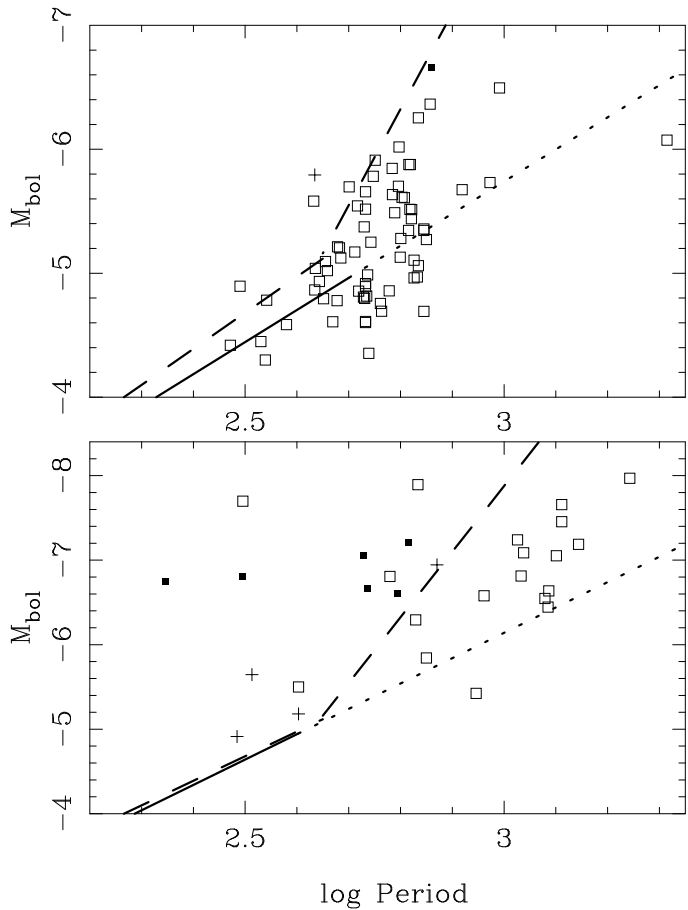}}
\end{minipage}

\caption[]{ 
As Fig.~\ref{Fig-MbolPer} but without the model tracks, without
supergiants, and only variables with large amplitudes are plotted.
Open squares indicate stars without information on the presence of
HBB, plus-signs are stars without HBB (no Lithium detected), filled
squares stars which undergo HBB (or supposed to, like the C-star 
IRAS 04496-6958, but which probably is not, see text).
}
\label{Fig-MbolPerHBB} 
\end{figure}

Figure~\ref{Fig-MbolPerHBB} shows the data in a different light, only
plotting the large amplitude (Mira) variables, and highlighting the
knowledge regarding the presence of Lithium, an indicator of HBB (data
from Smith \& Lambert 1989, 1990, Smith et al. 1995). It shows the
well known result that the known Li-rich stars are O-rich stars
located in the interval $-7.2 \less M_{\rm bol} \less -6.0$ (see the
discussion and Fig.~6 in Smith et al. 1995). The C-star marked as
undergoing HBB is IRAS 04496-6958 which was suggested to be in that
state by Trams et al. (1999).  At that time ISO spectra suggested a
silicate dust shell, although the central star is C-rich.  They
suggested that the star had just become a C-star, after HBB had
ceased. The fact that it had the largest deviation from the
$PL$-relation in the sample studied by Whitelock et al. (2003) seemed
to support this hypothesis as this is also observed in O-rich stars.
The much higher quality IRS spectrum shows no evidence for the
presence of a silicate feature, as was originally discussed by
  Speck et al. (2006), who suggested that this apparent silicate
  emission is an artifact of the underestimation of the level of the
  continuum emission in the ISO spectrum.  
In fact, there is no evidence that this star is currently undergoing HBB.

In the Whitelock et al. sample it was the only C-star that lay
significantly above the $PL$-relation, but one of the strengths of the
present study is that it enlarges significantly the number of heavy
mass-losing stars with a pulsation period.
Figure~\ref{Fig-MbolPerHBB} shows that IRAS 04496 still is the C-star
that has the largest overluminosity w.r.t. classical $PL$-relations
derived for shorter period miras, but no longer the only one. The
relation proposed by Hughes \& Wood (1990) seems to delineate the
upper boundary of luminosities that can be reached. For periods longer
than $P$= 450 days, where Hughes \& Wood proposed a break in the
$PL$-relation, there is enormous scatter, which is not present in the
$PL$-relation at shorter periods (the scatter found by Groenewegen \&
Whitelock is 0.26 mag). This may be in part due to the fact that the
luminosities in the present study are not from single-epoch photometry.

One peculiar object is MSX LMC 775, which is plotted at the period
with the largest amplitude, 2063 days as derived from OGLE data, 2210
days as derived from MACHO red data. This is very unusual as the
periods of C-stars are confined to $\less$1000 days (see Figure~\ref{Fig-MbolPerHBB}, 
Whitelock et al. 2003, Kerschbaum et al. 2006 for Galactic C-stars),
and this is consistent with the picture that HBB prevents the
formation of C-stars for large initial masses.  This star could truly
be an object that has turned into a C-star {\it after} HBB ceased.  On
the other hand this star also shows a significant period of 269 days
with an amplitude of 0.46 in the $I$-band, also suggestive of mira
pulsation. This would place the star in the upper left part in
Figure~\ref{Fig-MbolPerHBB}, also in a location not occupied by other C-stars. 
From the MACHO data a third period of 549 days can be derived, which
would place it among other C-stars, but the amplitude is smaller and
unlike the pulsation of a Mira.

Regarding the O-rich Miras there are a few that follow the
extrapolation of the classical $PL$-relation derived for short-period
miras, but most are overluminous. Some are known to be rich in Lithium
(the filled symbols) and it would be interesting to investigate this
for all objects in Figure~\ref{Fig-MbolPerHBB} brighter than $-6.0$ in
bolometric magnitude (although for some this will be difficult as they
are very red).





\section{Summary and conclusions}


Mass-loss rates for the sample are derived under the assumption of a
constant expansion velocity and dust-to-gas ratio with values similar
to galactic AGB stars.

Wachter et al. (2008) compute dynamical wind models for subsolar
metallicities for carbon stars.  They find that that the outflow
velocities of the solar metallicity models are higher by about a
factor of 2.2 $\pm$ 0.2 than those of the LMC, and 4 $\pm$ 1
than those of the SMC. The dust-to-has ratios are larger by factors
1.3 $\pm$ 0.1 and 2.3 $\pm$ 0.2, respectively.  For the modelling in
this paper this means that the assumption of typical velocities and
dust-to-gas ratios of galactic stars may lead to an overestimate of
the mass-loss rates by a factor of 1.7 $\pm$ 0.2 for the LMC,
and 1.7 $\pm$ 0.5 for the SMC targets, respectively (since the dust
optical depth $\tau \sim $\mdot$ \Psi / v_{\rm exp}$ is fitted).  
From their Eqs.~1-3 for default parameter values ($M$= 1\msol, 
$T_{\rm eff}$= 2600 K, $L$= 10$^4$\lsol), the ratio of mass-loss rates is 
GAL : LMC : SMC= 1: 0.84 : 0.52.

Mattsson et al. (2009) computed a grid of 900 dynamical model
atmospheres for carbon stars for solar metallicities spanning a range
in luminosity, pulsation velocity amplitude, effective temperature,
(C-O) excess and stellar mass. The online material includes output
(mass-loss rates, expansion velocities and dust-to-gas ratios) for
about 325 models.  Taking only models with expansion velocities in the
range 2-45 \ks\ only 253 ``plausible'' models remain. The 98\%
percentile on $\beta$ = (\mdot $v_{\rm exp})/(L/c)$ = 3.0.
Given the scaling parameters as derived by Wachter et al. one would
expect maximum values for $\beta$ a factor of 2.6 lower in the LMC and
7.7 lower in the SMC, respectively, than in our galaxy, or about 1.1 and 0.39.
From Table~\ref{Tab-Cstar} the 98\% percentile on $\beta$ = 1.2, while the
largest value for an SMC C-star is 0.52. These values are consistent with predictions.
The range in dust-to-gas ratio in the 253 models by Mattsson et al. is (3-80) 10$^{-4}$,
which encompasses the adopted value of 0.005.
  
At the level of a factor 2-4, the predicted and observed mass-loss rates
agree and confirm that there is no strong dependence of mass-loss rate on metallicity.
To have a better understanding of the mass-loss process it is crucial
to measure the expansion velocity and determine the gas-to-dust ratio
by observing the CO thermal emission lines in the (sub-)mm. In
addition it would be valuable if the C/O ratio could be determined for
some stars in the sample, as the wind properties are predicted to
depend on the excess of carbon over oxygen atoms. \bigskip

\noindent
In this last paragraph of the paper we want again to draw the attention to
the pulsation properties of two remarkable objects (see Sect 5.5). 
The C-star MSX LMX 775, 
which we classified as having a pulsation period of over 2000 days.
This is much longer than the maximum period of all known Galactic and MC 
C-stars which is near 900 days.  This period is located however on an extension
of the C-star $PL$-relation derived for unobscured stars.  If true,
MSX LMC 775 could be a very massive star that has become a C-star \it after \rm HBB ceased.
The other remarkable object is the O-rich star MSX SMC 055 with a
period of 1749 days and at $M_{\rm bol}= -8.0$.  Its pulsation
amplitude and colour are consistent with that of a (mass-losing) Mira,
and not an RSG star.  Its luminosity is consistent with that predicted
for super-AGB stars, and therefore we propose this object as the most
likely super-AGB star candidate from an observational point of view.

\acknowledgements{  
This publication is partly based on the OGLE observations obtained
with the Warsaw Telescope at the Las Campanas Observatory, Chile,
operated by the Carnegie Institution of Washington.

This paper utilizes public domain data originally obtained by the
MACHO Project, whose work was performed under the joint auspices of
the U.S. Department of Energy, National Nuclear Security
Administration by the University of California, Lawrence Livermore
National Laboratory under contract No. W-7405-Eng-48, the National
Science Foundation through the Center for Particle Astrophysics of the
University of California under cooperative agreement AST-8809616, and
the Mount Stromlo and Siding Spring Observatory, part of the
Australian National University.

This research has made use of the SIMBAD database, operated at CDS, Strasbourg, France. 

The authors would like to thank Dr. Mikako Matsuura for making the
lists of O- and C-stars discussed in Matsuura et al. (2009) available.

MG would like to thanks Ariane Lan\c con and Eric Josselin for organising the 
``Intermediate Mass Stars $\leftrightarrow$ Massive Stars, A workshop
around causes and consequences of differing evolutionary paths'' meeting in 
Strasbourg, which was inspiring and resulted in some ideas being explored in this paper.
}

{} 
 
\begin{appendix}
\section{Multi-periodic AGB stars}

Some of the stars for which we (re-)analysed ASAS, MACHO and OGLE data
show multiple periods.  Table~\ref{Tab-Csample} and \ref{Tab-Msample}
list the (adopted) pulsation period. In the table below all derived
periods are given
The first period is the adopted pulsation period, then the other
periods are listed. Typically only one additional period is fitted,
because of the irregular behaviour. Only in exceptional cases a third
period is fitted.  Formal errors in the periods are also given. In the
case of MACHO data, the blue channel was analysed {\it only} if the
red channel data was corrupt.


There are  41 unique stars listed in the Table. 
The second period fitted is in the range 0.4-0.6 times the primary period in 10 cases. 
The amplitude of the second period is in the range 0.1-1.0 times the amplitude of the primary period.
These cases probably represent the situation where a pulsation mode and the next higher overtone mode are detected.
When there is an LSP, it is mostly in the range 3-10 times the primary period.

\begin{table*}
\caption{Multi-period information}

\begin{tabular}{lrrrrrrl}
\hline
Identifier &  Period & Amp. & Period & Amp. & Period & Amp.  & Remarks \\
\hline

MSX LMC 587  & 644 $\pm$ 2 & 0.32 & 340 $\pm$ 1.3 & 0.18 &  & & ASAS I \\
HD 271832    & 514 $\pm$ 1.1 & 0.16 & 52.66  $\pm$ 0.02 & 0.057 & 142.0 $\pm$ 0.2 & 0.060 & ASAS V \\

WBP 51       & 382.06 $\pm$ 0.04 & 0.11 &  439.80 $\pm$ 0.07 & 0.09 & 258.36 $\pm$ 0.06 & 0.03 & MACHO R \\
MSX LMC 791  & 436.31 $\pm$ 0.02 & 0.17 & 1654.1  $\pm$ 0.2  & 0.19 &  & & MACHO B \\
MSX LMC 1318 & 683.36 $\pm$ 0.07 & 0.14 &  303.94 $\pm$ 0.02 & 0.11 &  & & MACHO R \\
WOH S 264    & 453.44 $\pm$ 0.04 & 0.10 & 2830    $\pm$ 1    & 0.29 &  & & MACHO B \\
MSX LMC 1492 & 559.00 $\pm$ 0.04 & 0.71 & 2309.4  $\pm$ 0.4  & 1.34 &  & & MACHO R \\
MSX LMC 1492 & 564.29 $\pm$ 0.12 & 0.59 & 2545.1  $\pm$ 1.4  & 1.06 & 276.24 $\pm$ 0.04 & 0.30 & OGLE I \\
MSX LMC  218 & 670.5  $\pm$ 0.2  & 0.52 & 2204    $\pm$ 4    & 0.26 &  & & MACHO R \\
MSX LMC  218 & 662.8  $\pm$ 0.2  & 0.74 & 3008    $\pm$ 7    & 0.36 &  & &  OGLE I \\
MSX LMC  775 & 2209.1 $\pm$ 0.85 & 0.67 & 281.99  $\pm$ 0.03 & 0.34 & 548.7 $\pm$ 0.13 & 0.27 & MACHO R \\
MSX LMC  775 & 2063   $\pm$ 2.3  & 1.83 & 269.46  $\pm$ 0.05 & 0.46 &  &  & OGLE I \\
WOH G 64     & 855.79 $\pm$ 0.03 & 0.68 & 2647    $\pm$ 1.0  & 0.26 &  & & MACHO R \\
HV 11366     & 182.787 $\pm$ 0.003 & 0.32 & 292.02 $\pm$ 0.02 & 0.14 &  & & MACHO B \\

MSX LMC 768  & 626.7 $\pm$ 0.5 & 0.91 & 3501 $\pm$ 27 & 0.64 & 317.6 $\pm$ 0.7 & 0.15 & OGLE I \\
MSX LMC 1282 & 655.1 $\pm$ 0.58 & 0.94 & 5503 $\pm$ 68 & 1.80 &  & & OGLE I \\
MSX LMC  937 & 658.9 $\pm$ 1.0 & 0.69 & 1515 $\pm$ 26 & 0.13 &  & & OGLE I \\
MSX LMC 1205 & 558.5 $\pm$ 0.2 & 0.94 & 3015 $\pm$ 7 & 1.20 &  & & OGLE I \\
MSX LMC  663 & 422.0 $\pm$ 0.5 & 0.02 & 3520 $\pm$ 18 & 0.05 &  & & OGLE I \\
MSX LMC  220 & 624.99 $\pm$ 0.08 & 0.58 & 2305.3 $\pm$ 0.9 & 0.62  &  & & OGLE I \\
MSX LMC 1308 & 501.9 $\pm$ 0.2 & 0.71 & 3430 $\pm$ 22 & 4.22 &  & & OGLE I \\
MSX LMC   95 & 609.06 $\pm$ 0.10 & 1.05 & 1664 $\pm$ 2  & 0.28 &  & & OGLE I \\
MSX LMC 1120 & 635.28 $\pm$ 0.21 & 0.89 & 1835 $\pm$ 2 & 0.79 & & & OGLE I \\

MSX SMC 066  & 521.00 $\pm$ 0.05 & 0.36 & 1837.3 $\pm$ 0.2 & 0.96 &  & & OGLE I \\
MSX LMC  438 & 615.0 $\pm$ 0.6 & 0.64 & 1864 $\pm$ 2 & 1.24 &  & & OGLE I \\
MSX SMC 014  & 317.0 $\pm$ 0.7 & 1.17 & 2811 $\pm$ 56 & 1.56 &  & & OGLE I \\
NGC 419 IR1  & 476.97 $\pm$ 0.04 & 0.77 & 4778 $\pm$ 5 & 0.59 &  & & OGLE I \\
MSX LMC  783 & 515.1 $\pm$ 0.2 & 0.85 & 2206 $\pm$ 10 & 0.35 &  & & OGLE I \\
MSX LMC  634 & 484.42 $\pm$ 0.10 & 0.77 & 1463.3 $\pm$ 1.1 & 0.51 &  & & OGLE I \\

ISO 00549    & 683.3 $\pm$ 0.2 & 0.56 & 3029.2 $\pm$ 1.4 & 1.10 &  & & OGLE I \\

ISO 00548    & 432.54 $\pm$ 0.03 & 1.11 & 1345.8 $\pm$ 0.4 & 0.42 &  & & OGLE I \\ 

MSX SMC 093  & 457.82 $\pm$ 0.l8 & 0.28 & 3159 $\pm$ 3 & 0.97 &  & & OGLE I \\
MSX LMC  754 & 448.16 $\pm$ 0.10 & 0.71 & 2303 $\pm$ 10 & 0.23 &  & & OGLE I \\ 
MSX SMC 232  & 466.73 $\pm$ 0.13 & 0.75 & 2452 $\pm$ 2 & 1.44 &  & & OGLE I \\
NGC 419 LE16 & 423.25 $\pm$ 0.04 & 0.29 & 1882.4 $\pm$ 0.3 & 0.64 &  & & OGLE I \\
MSX SMC 055  & 1749.1 $\pm$ 0.2 & 0.87 & 901.9 $\pm$ 0.2 & 0.24 &  & & OGLE I \\
IRAS 04509-6922 & 1240.70 $\pm$ 0.11 & 1.50 & 659.32 $\pm$ 0.11 & 0.41 &  & & OGLE I \\
IRAS 04516-6902 & 1084.17 $\pm$ 0.14 & 1.32 & 5436 $\pm$ 31 & 0.65 &  & & OGLE I \\
MSX LMC  642 & 1122.46 $\pm$ 0.13 & 1.65 & 552.1 $\pm$ 0.2 & 0.24 &  & & OGLE I \\
IRAS 05558-7000 & 1176.6 $\pm$ 0.3 & 2.34 & 556.5 $\pm$ 0.3 & 0.34 & & & OGLE I \\

MSX SMC 024  & 417.97 $\pm$ 0.11 & 0.12 & 227.40 $\pm$ 0.05 & 0.08 & & & OGLE I \\
IRAS 05003-6712 & 909.38 $\pm$ 0.11 & 1.39 & 446.68 $\pm$ 0.14 & 0.29 & & & OGLE I \\
MSX SMC 134  & 247.90 $\pm$ 0.12 & 0.042 & 140.75 $\pm$ 0.04 & 0.04 & & & OGLE I \\
WBP 77       & 215.21 $\pm$ 0.01 & 0.37 & 108.025 $\pm$ 0.008 & 0.12 & & & OGLE I \\


\hline
\end{tabular}
\end{table*}

\end{appendix}

\end{document}